\documentclass[%
 reprint,
 superscriptaddress,
 amsmath,amssymb,
 prl
]{revtex4-2}

\usepackage{graphicx}% Include figure files
\usepackage{siunitx}
\usepackage{amsfonts}
\usepackage{mathtools}
\usepackage{gensymb}
\usepackage{bm}
\usepackage{amsmath}
\usepackage[colorlinks=true,bookmarks=false,citecolor=blue,linkcolor=blue,hyperfootnotes=true,urlcolor=blue]{hyperref}
\usepackage{nicefrac}
\usepackage{comment}

\usepackage{lineno}

\usepackage{soul}        % for \hl
\usepackage{xcolor}      % for setting highlight color
\usepackage{etoolbox}    % for conditional checking
\sethlcolor{yellow!0}   % optional: customize highlight color

\DeclareRobustCommand{\new}[1]{\hl{#1}}

\begin{document} 
%\linenumbers
%\title{Slow dynamic phase fluctuations as a route to short-range order in superconducting Cu$_x$TiSe$_2$}

%Previous title: Impurity stabilized charge density wave fluctuation in layered dichalcogenides

\title{%Metastable short-range charge order from slow phase dynamics in superconducting Cu$_x$TiSe2
%Metastable short-range charge order in superconducting Cu$_x$TiSe$_2$
Metastable domains from quenched disorder in superconducting Cu$_x$TiSe$_2$}

\author{Thomas Sutter}
\affiliation{Department of Physics and Astronomy, University of California Los Angeles, Los Angeles, CA}

\author{Colleen Lindenau}
%\email{xmchen@bnl.gov}
\affiliation{Department of Physics, Drexel University, Philadelphia, PA}

\author{Shivani Sharma}
%\email{xmchen@bnl.gov}
\affiliation{Brookhaven National Laboratory, Upton, NY}

\author{Andrei Fluerasu}
%\email{xmchen@bnl.gov}
\affiliation{Brookhaven National Laboratory, Upton, NY}

\author{Lutz Wiegart}
%\email{xmchen@bnl.gov}
\affiliation{Brookhaven National Laboratory, Upton, NY}

\author{Goran Karapetrov}
%\email{xmchen@bnl.gov}
\affiliation{Department of Physics, Drexel University, Philadelphia, PA}

\author{Anshul Kogar}
\email{anshulkogar@physics.ucla.edu}
\affiliation{Department of Physics and Astronomy, University of California Los Angeles, Los Angeles, CA}

\author{Xiaoqian M Chen}
\email{xmchen@bnl.gov}
\affiliation{Brookhaven National Laboratory, Upton, NY}

%%% REVISION #1 %%%

\begin{abstract}
In a vast array of materials, including cuprates, transition metal dichalcogenides (TMDs) and rare earth tritellurides, superconductivity is found in the vicinity of short-range charge density wave (CDW) order. The crossover from long-range to short-range charge order often occurs as quenched disorder is introduced, yet it is unclear how this disorder disrupts the CDW. Here, using x-ray photon correlation spectroscopy (XPCS), we investigate the prototypical TMD superconductor Cu$_x$TiSe$_2$ and show that disorder induces substantial CDW dynamics. We observe CDW phase fluctuations on a timescale of minutes to hours above the nominal transition temperature while the order parameter amplitude remains finite. These long timescale fluctuations prevent the system from finding the global free energy minimum upon cooling and ultimately traps it in a short-range ordered metastable state. Our findings demonstrate how correlated disorder can give rise to a distinct mechanism of domain formation that may be advantageous to the emergence of superconductivity.
\end{abstract}
\maketitle

\begin{figure}[t!]
    \includegraphics[width=1\linewidth]{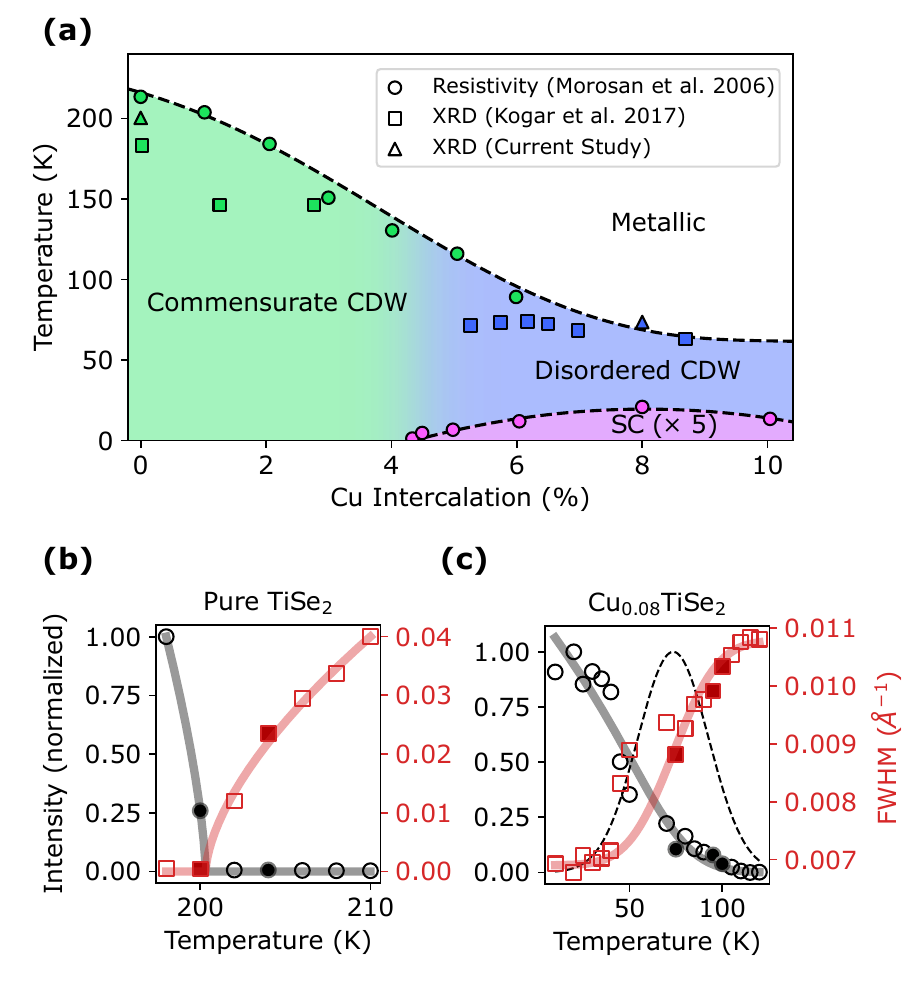}
    \caption{\textbf{CDW Phase Evolution in 1\textit{T}-Cu$\mathbf{_x}$TiSe$\mathbf{_2}$} (a) Temperature–copper fraction phase diagram illustrating the crossover from a commensurate CDW phase to a disordered CDW phase, together with the superconducting dome (SC). Data points are taken from \citet{Morosan_Cava_2006, Kogar_Rosenkranz_2017}. (b,c) Temperature dependence of the normalized XRD intensity and peak width of the $(0.5, 0.5, \overline{4.5})$ CDW reflection for the pure and 8\% intercalated samples, respectively. The data are fitted with a power-law curve (red) and a Gaussian-smeared power-law curve (gray), respectively. The black dashed line marks the width of the Gaussian used for the convolution. XPCS results are displayed for filled data points in Fig.~\ref{fig:pure_xpcs} and Fig.~\ref{fig:doped_xpcs}.
    \label{fig:order_param}}
\end{figure}

%\subsection{Introduction}
Unconventional superconductors, which do not conform to the Bardeen-Cooper-Schrieffer (BCS) paradigm, are often found as a neighboring phase is suppressed by doping, disorder, or applied pressure. 
%Superconductivity resides in a dome-like shape next to spin-density wave order in iron pnictides \cite{Cai_Yayu_2013, Luetkens_Buchner_2009, Kamihara_Hideo_2008, Takahashi_Hideo_2008}, antiferromagnetism in cuprates \cite{Rybicki_Jurgen_2016, Mello_Dias_2007, Shen_Seamus_2008}, ``hidden order" in URu$_2$Si$_2$ \cite{Mydosh_Riseborough_2020}, nematic order in BaNi$_2$As$_2$ \cite{Yao_Matthieu_2022}, and charge density wave order in transition metal dichalcogenides \cite{Shi_Yoshihiro_2015, Sipos_Tutis_2008} and rare-earth tritellurides \cite{Siddique_Cha_2024}. This universality suggests that superconductivity either competes with or cooperates with the nearby phase. 
Charge density wave  (CDW)-hosting transition metal dichalcogenides (TMDs) generally possess phase diagrams that are less complex than those of many other classes of unconventional superconductors. They may therefore serve as idealized platforms through which to investigate the relationship between superconductivity and a proximal phase. Although electrical transport studies of several TMDs were initially suggestive of a CDW quantum critical point within the superconducting dome, recent diffraction and scanning tunneling microscopy work indicate a more subtle relationship between the two phases \cite{Morosan_Cava_2006, Kusmartseva_Tutis_2009,  Sipos_Tutis_2008, Hinlopen_Friedemann_2024, Yu_2015}. In the canonical TMD system 1$T$-TiSe$_2$, CDW domain walls have been observed near the onset of the superconducting dome upon copper intercalation (Fig.~\ref{fig:order_param}~(a)), the application of hydrostatic pressure, and electric gating of thin films \cite{Yan_Madhavan_2017, Kogar_Rosenkranz_2017, Joe_Abbamonte_2014, Li_Neto_2015}. Remnant CDW order persists across the superconducting region, though only in short-range form  \cite{Novello_Renner_2017}. How these CDW textures develop and why they appear coincidentally with superconductivity is at present controversial, though an emerging view is that the short-range order is intimately tied to the physics of disorder~\cite{Novello_Renner_2017, Kogar_Rosenkranz_2017, Chatterjee_Rosenkranz_2015, Arguello_Pasupathy_2014, Yan_Madhavan_2017}.

From the perspective of mean field theory, disorder can couple to the order parameter, $\psi(r)$, through the various terms in the free energy \cite{Harris_1974, Imry_Ma_1997, Brock_Sweetland_1994, Vojta_Rastko_2004}:
\begin{align}
\label{eq:LGW}
\mathcal{F[\psi]} = \int d^d x \big[ -h&(\vec{x})\psi\left(\vec{x} \right) + (r + \delta r(\vec{x}))\psi^2\left(\vec{x}\right)\\
&+ u\psi^4(\vec{x}) + \cdots \big], \notag
\end{align}
where $r$ and $u$ are Landau expansion coefficients whereas $h$ and $\delta r$ are associated with field and $T_c$ (or temperature) disorder respectively. The former pins the phase of the CDW, while the latter is a spatial variation in the local transition temperature which does not bias the phase \cite{Vojta_2013}. In CDW systems, quenched disorder often refers to field disorder, which underlies effects like CDW pinning, sliding and narrow band noise~\cite{Gruner}. We show in this work that inhomogeneous CDW textures and their fluctuations in Cu$_x$TiSe$_2$ is instead, rather surprisingly, governed by correlated $T_c$ disorder. Our work highlights the importance of the comparatively unexplored effects of $T_c$ disorder in CDW systems, which we show may be advantageous for superconductivity.

%Our work highlights the importance of this comparatively unexplored T c disorder and its possible relevance to the emergence of superconductivity.

%To experimentally investigate the dynamics of the mesoscopic CDW textures, we use x-ray photon correlation spectroscopy (XPCS), which uses coherent x-rays to track both the amplitude and phase distribution of the CDW. Such information is encoded in an interference (or speckle) pattern within the diffraction peaks \cite{Zhang_Alec_2018}. By obtaining a time series of speckle patterns, XPCS provides a time resolved view of the CDW textures. %Our study reveals an uncommon mechanism of short-range CDW formation governed by quenched disorder, by demonstrating that the intercalated copper ions can introduce an inhomogeneous local CDW transition temperature in $1T$-TiSe$_2$ which fundamentally alters the time and correlation length of CDW fluctuations near the transition.
To investigate the dynamics of the mesoscopic CDW textures in Cu$_x$TiSe$_2$, we use x-ray photon correlation spectroscopy (XPCS), which uses coherent x-rays to track both the amplitude and phase distribution of the CDWs \cite{shpyrko2007direct, Chen_Wilkins_2016, chen2019charge, campi2022nanoscale, porter2024understanding, chen2019spontaneous}. The coherent x-ray scattering measurements were conducted at ID-11 at the NSLS-II Synchrotron light source. The probe x-ray beam is at 12.8~keV photon energy and is focused to a 5~$\mu$m spot-size on the sample. We examine the CDW fluctuations in the pristine ($x=0$) and optimally intercalated ($x=0.08$) compounds. The pristine sample possesses long-range 2$a\times$2$a\times$2$c$ CDW order, while only short-range charge order is present in the intercalated sample~\cite{Kogar_Rosenkranz_2017}. The amplitude and phase distribution of the CDW is encoded in the coherent x-ray interference (or speckle) pattern within the diffraction peaks \cite{Zhang_Alec_2018}. By obtaining a time series of speckle patterns, XPCS provides a time resolved view of the CDW textures. 

%\subsection{sample}

%\subsection{Results}

The temperature dependence of the (0.5, 0.5, $\overline{4.5}$) CDW peak intensity and width in both pristine and copper-intercalated 1$T$-TiSe$_2$ are plotted in Fig.~\ref{fig:order_param}(b)--(c).
In the pristine system, a resolution-limited CDW peak appears below $T_{\text{CDW}}=200$~K which is indicative of long-range order~\cite{Tc}. Only diffuse scattering is present above $T_{\text{CDW}}$. 
%The full width at half maximum (FWHM) of the diffuse scattering above $T_{\text{CDW}}$ is shown in Fig.~\ref{fig:order_param}(d). 
%Critical exponents of the intensity and the correlation length, $I (T{<}T_{\text{CDW}}) \sim (T_c-T)^{2\beta}$ and FWHM$(T{>}T_{\text{CDW}}) \sim (T-T_c)^{\nu}$, are consistent with the three-dimensional Ising universality class ($\beta \approx 0.33$ and $\nu\approx 0.63$) which is expected given that the phase of the individual components of the CDW can only take the values zero or $\pi$ \cite{Novello_Renner_2017, Spera_Renner_2019}.
%due to the commensurate nature of the CDW. %This is expected for a single dimerizing CDW because the phase is a binary scalar. In 1$T$-TiSe$_2$, the situation is somewhat more complex because there are three separate dimerizing CDWs that may couple to each other.
% Q1 changes in this para
For the copper-intercalated sample, the transition broadens significantly and the nominal transition temperature is suppressed to $T_{\text{CDW}}^{\text{Cu}}$=74~K (Fig.~\ref{fig:order_param}(c))~\cite{Tc}. %In Fig.~\ref{fig:order_param}(c),
%the CDW intensity is shown as a function of temperature. The fit line assumes that the transition temperature is smeared by disorder, which is modeled by a convolution with a Gaussian curve. 
%$T_{\text{CDW}}^{\text{Cu}} \approx 74$ K, while the transition temperature also broadens considerably \cite{Tc}. 
%\new{It should be noted that the distribution of local $T_c$ values across the sample is not necessarily Gaussian; the Gaussian form is used here solely as a convenient phenomenological fit to estimate the width of the disorder broadened transition.}
%%%
%The FWHM of the CDW peak in Cu$_{0.08}$TiSe$_2$ is shown as a function of temperature in Fig.~\ref{fig:order_param}(e). 
%The sample no longer exhibits long-range order, as the FWHM does not become resolution limited down to the lowest measured temperature of 10 K. At this temperature, the correlation length is roughly 140 $\textrm{\AA}$ or 40 unit cells. 
In contrast to the pristine sample, which exhibits a sharp transition from a resolution-limited peak to broad diffuse scattering at $T_{\text{CDW}}$, the copper-intercalated system shows a continuous evolution with a finite FWHM gradually broadening with increasing temperature.

\begin{figure}
\includegraphics[width=0.95\linewidth]{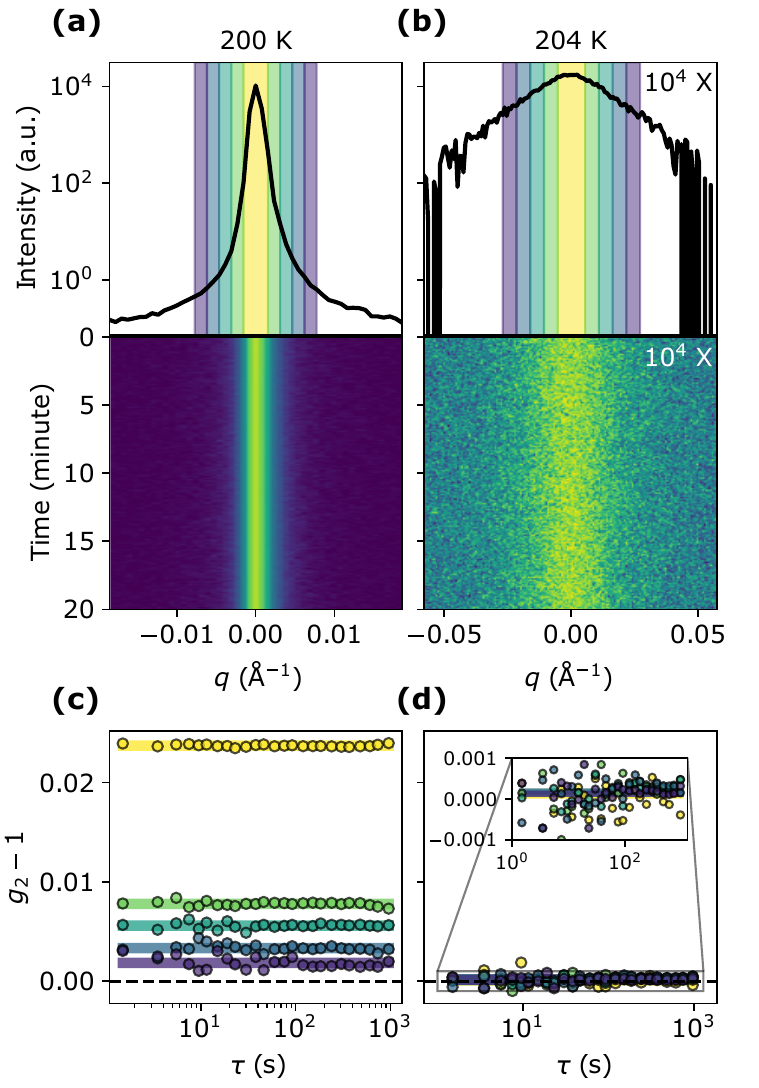}
    \caption{\textbf{XPCS on a CDW peak of pristine 1\textit{T}-TiSe$\mathbf{_2}$}. All data is in reference to the (0.5, 0.5, $\overline{4.5}$) peak. The 1st column shows 200 K data, and the 2nd column shows 204 K data. (a), (b), Line cuts along $h$ of the CDW peak on a log scale. The 204 K line cut in (b) is scaled by $10^4$. In the top panels, the shaded regions are different ROIs for XPCS, and the curves are average line cuts over all times. The bottom plots are log scale waterfall images in which the line cut at a specific frame is rendered along the horizontal axis while the vertical axis is time. The duration of a single frame is 1.0 s. (c), (d) Speckle correlation as a function of lag time. The colors of the curves indicate the associated ROIs shown in (a) and (b). The inset in sub-figure (d) is a magnification of the rectangular region indicated.}
    \label{fig:pure_xpcs}
\end{figure}

\new{It is evident that} the disorder introduced by copper intercalation significantly \new{alters the characteristics of the CDW transition} in 1$T$-TiSe$_2$.
Figure~\ref{fig:pure_xpcs} summarizes the XPCS results for the pristine system by depicting two temperature points with characteristic behaviors below and above $T_{\text{CDW}}$ (filled points in Fig.~\ref{fig:order_param}(b)). To visualize the dynamics, the time evolution of a line cut through the diffraction peak is displayed in the ``waterfall" plots (Fig.~\ref{fig:pure_xpcs}(a)--(b)). At all measured temperatures, we observed no time variation in the speckle \new{correlations} for an acquisition time of 20 minutes. These results indicate that the CDW is either static or fluctuates faster than the experimental time resolution of 10 ms, the shortest exposure time of the detector \cite{fast}.

\begin{figure*}[htb!]
    \includegraphics[width=0.89\linewidth]{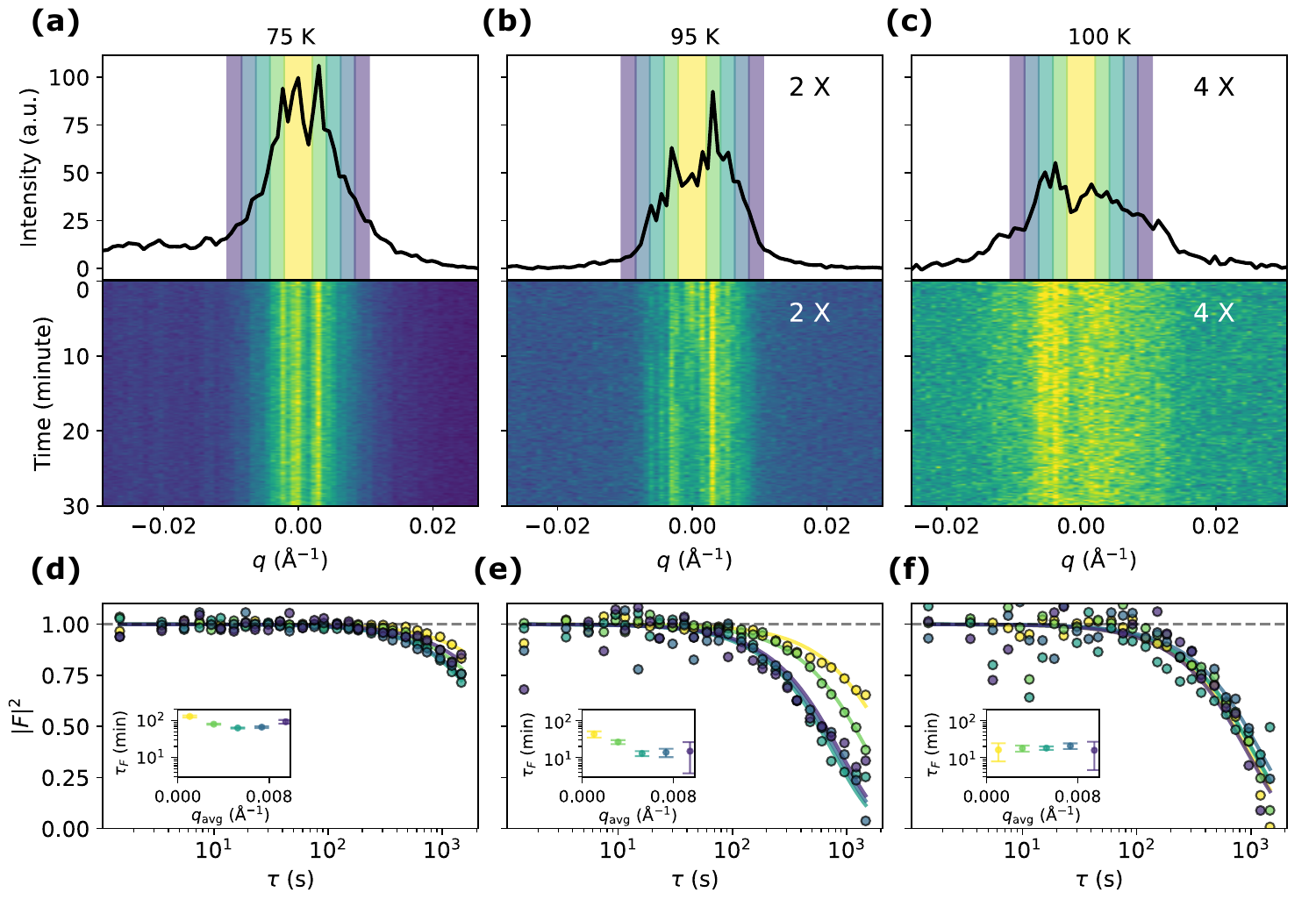}
    \caption{\textbf{XPCS on a CDW peak of Cu$_{\mathbf{0.08}}$TiSe$\mathbf{_2}$}. All data is in reference to the (0.5, 0.5, $\overline{4.5}$) CDW peak. The 1st, 2nd, and 3rd columns show data at 75 K, 95 K, and 100 K respectively. (a)--(c), Line cuts along $h$ of the CDW peak. In the top panels, the shaded regions indicate different ROIs for XPCS, and the curves are average line cuts over all times. The bottom plots are waterfall images in which the line cut at a specific frame is rendered along the horizontal axis while the vertical axis is time. The duration of a single frame is 1.0 s. (d)--(f) The intermediate scattering function plotted against lag time. The colors of the curves indicate the associated ROIs shown in (a)--(c). Trend lines are fits to the compressed exponential of $\exp\left({-\left(\frac{\tau}{\tau_F}\right)^{\gamma}}\right)$ with $\gamma$ = 1.2. The insets show the correlation decay times $\tau_F$ where the x-axis is the average reciprocal space distance from the Bragg peak centroid of the associated ROI.} 
    \label{fig:doped_xpcs}
\end{figure*}

The speckle time dependence was quantified using the normalized equal-time correlation of speckle intensity:
\begin{align}
\label{eq:g2}
g_2(\vec{q}, \tau) &= \frac{\langle I(\vec{q}, t)I(\vec{q},t + \tau) \rangle}{\langle I(\vec{q}, t) \rangle^2} \\
&= 1 + \beta_c \vert F(\vec{q}, \tau) \vert^2 \notag
\end{align}
where $I(\vec{q}, t)$ is the speckle intensity at momentum transfer $\vec{q}$. To improve the signal-to-noise ratio and to permit the assessment of static correlations, the brackets $\langle \rangle$ in Eq.~\ref{eq:g2} are taken to indicate both an average over all times and an average over all $\vec{q} \in \mathcal{Q}$, where $\mathcal{Q}$ is in practice a collection of detector pixels. $g_2$ is often expressed in terms of an intermediate scattering function $\vert F(\vec{q},\tau)\vert$. The parameter $\beta_c$ is the speckle contrast factor that ranges from 0 (no speckle contrast) to 1 (perfect speckle contrast) depending on the experimental setup~ \cite{Chen_Wilkins_2016}.

In Fig.~\ref{fig:pure_xpcs}~(c)--(d), we plot $(g_2(\tau)-1)$ calculated over five tightly packed annular regions of interest (ROIs) surrounding the centroid of the peak with outer radii increasing successively by 0.0015 $\textrm{\AA}^{-1}$ and 0.0053 $\textrm{\AA}^{-1}$ for the 200~K and 204~K data respectively. Below $T_\text{CDW}$, the measured $g_2$ is greater than one and time independent in all ROIs, indicating a static CDW texture (Fig.~\ref{fig:pure_xpcs}(c)). The decrease in speckle contrast, $\beta_c$, with increasing $q$ reflects the reduced CDW intensity at higher $q$ \cite{Chen_Wilkins_2016}.  Whereas $g_2(\tau)$ was greater than one below $T_c$, above $T_\text{CDW}$, it remains flat at unity for all ROIs (Fig.~\ref{fig:pure_xpcs}(d)). This complete loss of speckle contrast indicates that above $T_\text{CDW}$, the scattering feature arises from dynamic CDW fluctuations.
Overall, the data on the pristine system are consistent with expectations for a continuous phase transition. 
%A regime of diverging fluctuation timescales, as predicted by critical slowing down, is expected to occur within a very narrow temperature window that was not accessed in this study. 
%These results provide a useful reference for comparison with the XPCS data on the copper intercalated system, where disorder plays a significant role in modifying the dynamics.
\begin{figure*}
    \includegraphics[width=1\linewidth]{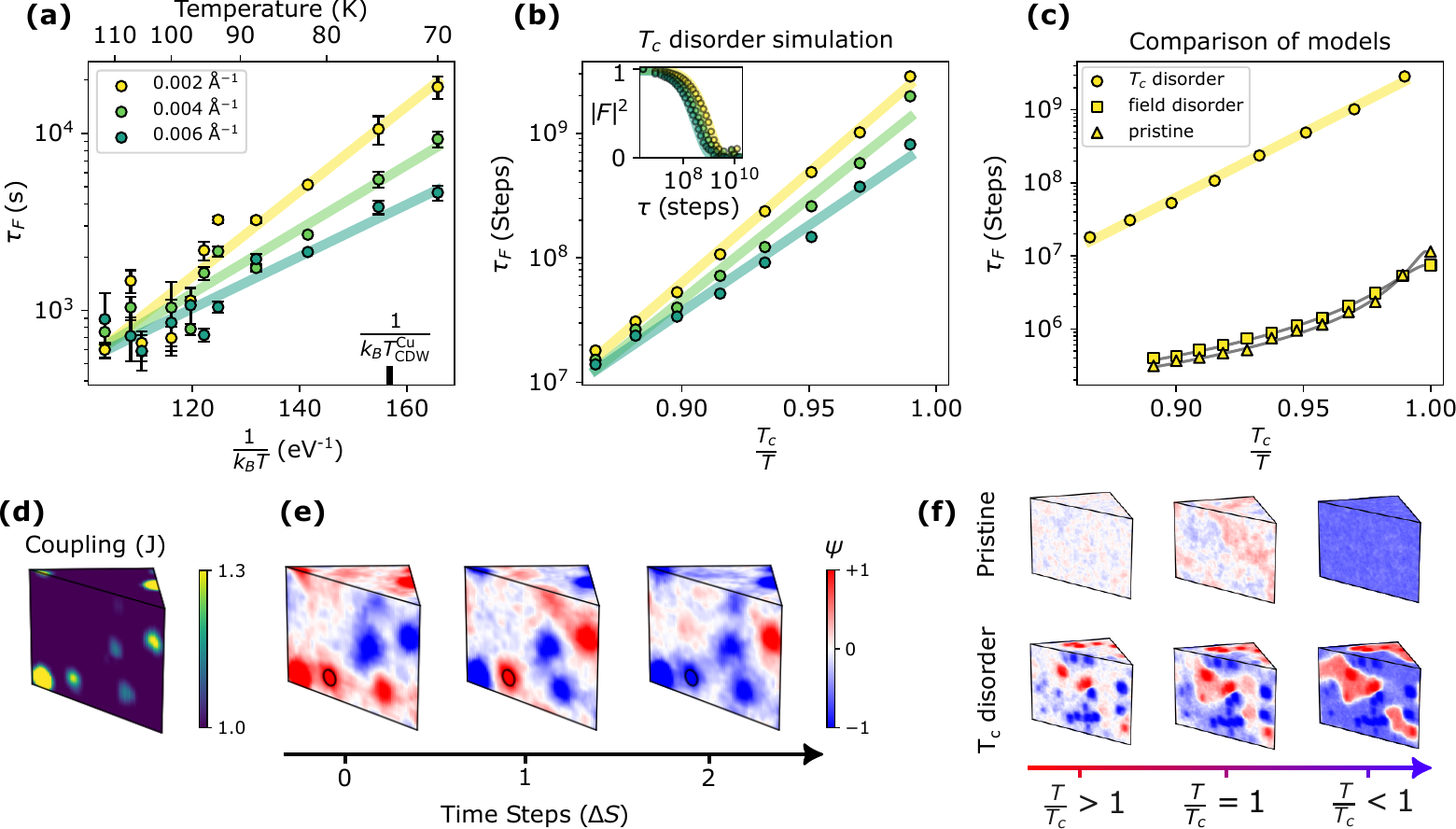}
    \caption{\textbf{Summary of XPCS data and simulations} (a) Fit value of the decay time constant $\tau_F$ as a function of inverse temperature for XPCS scans on the (${0.5}$, ${0.5}$, $\overline{4.5}$) CDW of the copper-intercalated sample. The error bars are from the covariance of the least squares fit. The legend indicates the outer radius of the annular ROI regions for the data. The $\tau_F(T)$ data for each ROI are fit to the Arrhenius form. (b) $\tau_F$ as a function of normalized temperature for a correlated $T_c$ disorder simulation. The ROIs follow the same trend of increasing $q$ as in (a). The inset shows the time-dependence of the intermediate scattering function for the simulation temperature $T_c/T=0.96$. (c) Comparison of $\tau_F$ for $T_c$ disorder, field disorder, and pristine simulations. The lowest $q$ ROI is used for this comparison. (d) Map of the coupling parameter on a cut of the simulation grid for the $T_c$ disorder simulation (Disorder in $J$, the coupling strength, is used as a proxy for $T_c$ disorder \cite{Tc}). (e) Averaged snapshots of simulation states from the $T_c$ disorder simulation at $T_c/T=0.96$ where $\Delta S = 3 \times 10^9$ elementary simulation steps. (f) Comparison of pristine and $T_c$ disorder for a linear ramp cooling simulation. The simulation starts at $T/T_c = 1.04$ and cools to 0.96. The cooling rate is $\Delta T/T_c = -2 \times 10^{-12}$ per step.}
    \label{fig:dynamics}
\end{figure*}

In contrast to the pristine system, Cu$_x$TiSe$_2$ exhibits slow dynamics in the transition region. Figure \ref{fig:doped_xpcs} summarizes XPCS data from Cu$_{0.08}$TiSe$_2$ at three temperatures close to and above the nominal 74 K transition temperature: 75 K, 95 K, and 100 K (filled points in Fig.~\ref{fig:order_param}~(c)). The time variation of the speckles reveal a slowing of the dynamics with decreasing temperature, which is quantified in Fig.~\ref{fig:doped_xpcs}~(d)--(f) through the intermediate scattering function (Eq. (2)). The reciprocal space ROIs are annular regions centered on the peak with outer radii increasing by 0.002 $\textrm{\AA}^{-1}$. At 100~K, $|F(\tau)|^2$ is roughly independent of $q$; the five different ROIs all decay on the same timescale (Fig.~\ref{fig:doped_xpcs}(f)). At 95~K and 75~K, the decay time increases and becomes $q$ dependent such that the lower $q$ ROIs show a longer timescale (Fig.~\ref{fig:doped_xpcs}(d)--(e)). 
%\new{Separate plots of $|F(\tau)|^2$ for individual ROIs are shown in Supplementary Fig.~2. The two-dimensional diffraction images for the peak profiles in Fig.~{\ref{fig:pure_xpcs}} and Fig.~{\ref{fig:doped_xpcs}} are provided in Supplementary Fig.~3}
XPCS data was also collected on the (0, 0, 2) Bragg peak of the copper-intercalated sample, which showed no dynamics even up to 300 K~\cite{Tc}. Thus, structural dynamics associated with mobility of the Cu intercalants is absent.

To quantify these measurements, we compute a correlation decay time on each XPCS dataset by fitting $|F(\tau)|^2$ to a stretched/compressed exponential:
$\lvert F (\tau) \rvert^2 = \exp\left({-\left(\frac{\tau}{\tau_F}\right)^{\gamma}}\right)$,
where $\tau_F$ and $\gamma$ are the decay time constant and stretching exponent respectively. For all the XPCS data, the $\gamma$ parameter of the fit shows no temperature or $q$ dependence, and it is fixed for all fits at the mean value of $\gamma = 1.2$ \cite{Tc}. The temperature dependence of $\tau_F$ for the three lowest-$q$ ROIs is shown in Fig.~\ref{fig:dynamics}(a). For each ROI, the temperature dependence of $\tau_F$ fits to an Arrhenius form,
$\tau_F(T) = \tau_0\exp\left(\frac{\Delta E}{k_BT}\right)$,
which is characteristic of thermally activated processes. The extracted activation energies are 55~meV, 42~meV, and 34~meV for $q = 0.002$~Å$^{-1}$, $0.004$~Å$^{-1}$, and $0.006$~Å$^{-1}$, respectively. \new{The corresponding prefactors are $\tau_0 = 2.03$ s, 8.37 s, and 17.90 s. The increasing trend of $\tau_0$ with $q$ is mathematically required to produce the observed convergence of the $\tau_F$ curves at approximately 110 K. %This loss of $q$ dependence suggests the dynamics at high temperature are associated with a single relevant length scale. %These $\tau_0$ values do not have a simple microscopic interpretation, as the CDW order essentially vanishes above 110 K and the infinite temperature limit is not accessible. However, the increasing trend of $\tau_0$ with $q$ is mathematically required to produce the observed convergence of the $\tau_F$ curves at approximately 110 K. This loss of $q$ dependence suggests the dynamics at high temperature are associated with a single relevant length scale, which we show below is related to the quenched disorder (See Supplementary Section II).
}

To understand these observations and their relationship to the underlying disorder, the speckle dynamics were simulated using a 3D disordered Ising model. The mapping of the CDW order to the Ising model is permitted because each of the three CDWs in TiSe$_2$ is effectively independent and possesses a $Z_2$ order parameter~\cite{guo2025plane, Tc}. We perform three simulations to demonstrate the difference between pristine, field disordered, and $T_c$ disordered phase transitions. 
%The field disorder is mediated through a site-dependent term $h_{ijk}$ (Eq.~\ref{eq:LGW}) that couples linearly to the order parameter $\psi_{ijk}$, while $T_c$ disorder is accomplished through a variable nearest-neighbor coupling $J_{ijk}$. Here, the $ijk$ are indices denoting site position on the simulated three dimensional lattice. 
To correspond with the scanning tunneling microscopy results of Ref.\cite{Spera_Renner_2019} on Cu$_x$TiSe$_2$, the generated disorder distributions have Gaussian spatial correlation, i.e. the disorder is \textit{correlated}. 
\new{The disorder distribution 
%of $J_{ijk}$ 
for the $T_c$ disorder simulation is illustrated in Fig.~{\ref{fig:dynamics}(d)}.}
The simulation is time evolved through Glauber dynamics, and the Fast Fourier Transform (FFT) of the system state is used to record simulated diffraction patterns. A $g_2$ correlation function can then be computed from these FFT snapshots \cite{Tc}.

A series of decay time constants, $\tau_F(T)$, obtained from the simulation of $T_c$ disorder, is illustrated in Fig.~\ref{fig:dynamics}(b). The simulated $\tau_F$ is expressed in elementary simulation steps, i.e., a single site update. The stretching exponent $\gamma$ is approximately one for this simulation irrespective of ROI or temperature \cite{Tc}.
The $T_c$ disorder simulation reproduces the exponential dependence of $\tau_F$ on inverse temperature along with the trend in $q$ observed in the data (Fig.~\ref{fig:dynamics}(a) and (b)). In contrast, the pristine and field disorder models produce $\tau_F(T)$ curves corresponding to a power law divergence due to critical slowing down. We can further distinguish the signatures of $T_c$ disorder and field disorder by comparing their fluctuation timescales to that of the pristine model. The field disorder model shows a suppressed timescale near the transition while the $T_c$ disorder model massively increases the timescale. A comparison of the three models ($T_c$ disorder, field disorder, and pristine) is illustrated in Fig.~\ref{fig:dynamics}(c). For simplicity, only the $\tau_F$ curves from the lowest $q$ ROI are shown \cite{Tc}.

For the $T_c$ disorder model, the dynamics are dominated by phase flipping of ordered droplets. This process is illustrated in Fig.~\ref{fig:dynamics}(e), where the regions of increased $T_c$ (modeled by disorder in the coupling strength, $J$) develop order above the nominal transition temperature but execute $\psi \rightarrow -\psi$ flips. %The time dependence of this phase flipping within a single droplet resembles a random telegraph signal. 
As the temperature is reduced and the free energy well gets deeper, surmounting the energy barrier becomes less likely which leads to a longer timescale. These timescales correspond to the thermally activated process that ensures $\tau_F(T)$ will follow an Arrhenius form \cite{Brendel_Henk_2003, Ricci_Buchholz_2021}. This massive increase in timescales, along with the exponential dependence on inverse temperature, can be reproduced by the $T_c$ disorder simulation but not by the field disorder simulation.

Taken together, the experimental data and Monte Carlo simulations support a picture in which thermally activated phase flipping, driven by correlated $T_c$ disorder, dominates the observed dynamics. The Arrhenius temperature dependence of $\tau_F$, the finite-temperature convergence of the different $\tau_F(T)$ curves (indicating a loss of $q$-dependence), and the emergence of long timescales all arise naturally within this framework. 
%Supplementary Table I further illustrates the agreement between simulation and experiment, showing that the model reproduces not only the trend of decreasing $\Delta E$ with increasing $q$, but also the overall magnitude of the activation energies. Together, these features highlight the central role of spatially varying local transition temperatures in producing the observed dynamics.}

The conclusion that copper intercalation induces correlated $T_c$ disorder, rather than field disorder, implies that the short-range ordered state at low temperatures emerges through an unconventional mechanism. The theoretical ground state of a $T_c$-disordered system is perfect, long-range order; however, correlated $T_c$ disorder precipitates the onset of metastable short-range order upon cooling through the transition. This metastability is illustrated in Fig.~\ref{fig:dynamics}(f) which compares the pristine and $T_c$ disorder models for the same cooling rate in which the system is initialized above $T_c$ and continuously cooled below \cite{Tc}. The cooling rate is slow enough that the pristine system achieves long range order; however, the $T_c$ disordered system develops domains seeded from the uncorrelated phases of the $T_c$ enhanced regions.

Because the disorder is correlated, a length scale is set by the average size and inter-island distance of the copper-rich and copper-poor regions as observed with STM \cite{Spera_Renner_2019}. This length scale ultimately determines the correlation length of the short-range order at low temperatures. Formation of metastable domains is therefore characterized by the following ingredients: (i) disorder that does not pin the phase of the order parameter, (ii) spatial correlation of the disorder, (iii) a phase coherence timescale that is much longer than an experimentally relevant cool-down time. As temperature decreases, the phase flipping time of the CDW islands gradually extends to hours. This exponentially increasing timescale associated with CDW phase coherence prevents the system from fully equilibrating, ultimately trapping it in a metastable short-range ordered state at low temperatures due to correlated temperature disorder. 

Recent STM results show that in Cu$_x$TiSe$_2$ there is no statistically significant correlation between the CDW phase and intercalant positions, which reinforces our view that copper intercalation introduces temperature disorder \cite{Spera_Renner_2019}. In contrast, when titanium atoms are intercalated, a strong correlation is observed, indicating field disorder \cite{Hildebrand_Aebi_2017}. Importantly, superconductivity is only observed upon copper intercalation, but not titanium intercalation.

Distinguishing between temperature disorder and field disorder may therefore be crucial to understanding how superconductivity emerges in CDW systems. Field disorder creates sharp domain walls that are rigidly pinned; on the other hand, domain walls induced by temperature disorder are much ``softer" precisely because of the lack of phase pinning. Even at low temperature, the boundaries of the temperature disordered regions will fluctuate, which can be advantageous for Cooper pairing. $T_c$ disorder is ultimately responsible for suppressing long-range order, allowing a finite density of states at the Fermi surface at the domain walls, and providing a fluctuating medium that instigates pairing. Notably, such considerations do not imply that superconductivity would be confined to the domain walls. Bulk superconductivity would still result if the superconducting healing length is longer than the characteristic domain size. Our results ultimately identify correlated temperature disorder as an organizing principle that reshapes the CDW transition in Cu$_x$TiSe$_2$. Such disorder favors metastable short-range order whose resulting domain structure naturally provides a setting in which superconductivity can emerge in close proximity to CDW textures.

\vspace{0.5em}
\begin{acknowledgments}
\emph{Acknowledgments}--- X-ray experiments were supported by the U.S. Department of Energy, Office of Basic Energy Sciences Grant No. DE-SC0012704(X.~C.), Grant No. DE-SC0023017 (A.~K.), and Grant No. ECCS-1711015(G.~K.)
\end{acknowledgments}

\bibliography{refs.bib}

\end{document}

\clearpage
\onecolumngrid
\renewcommand{\theequation}{S\arabic{equation}}
\renewcommand{\figurename}{\textbf{Supplementary Fig.}}

\renewcommand{\tablename}{\textbf{Supplementary Table}}

\renewcommand{\refname}{Supplementary References}  % For article class

\setcounter{figure}{0}
\setcounter{equation}{0}
\pagestyle{empty}

\tableofcontents

\newpage

\section{Supplementary Section I. Data analysis and fitting}
\label{sec:data-analysis}

The 2 $\times$ 2 $\times$ 2 commensurate CDW transition in $1T$-TiSe$_2$ is a critical transition; thus, the CDW amplitude at temperatures below $T_c$ should follow a power law scaling: $\langle a \rangle \propto (T_c-T)^{\beta_c}$ where the angled brackets denote an average over the entire system. The intensity of a CDW diffraction peak is proportional to the quantity $\langle a^2 \rangle$. This intensity is greater than zero even at temperatures above $T_c$ where $\langle a \rangle = 0$; such intensity is called diffuse scattering. At temperatures below $T_c$, $\langle a \rangle$ is much greater than the square-root of the variance in $a$; thus, we can associate the square-root of the intensity with the amplitude of the CDW. Consequently, one can fit the intensity of the CDW peak at temperatures below $T_c$ to the curve $(T_c-T)^{2\beta_c}$. In this work, diffraction data was only collected at two temperatures below $T_c$ for the pristine $1T$-TiSe$_2$ sample; this is insufficient data to fully fit the power law curve. However, the critical exponent has been measured previously as $2\beta_c = 0.68$ \cite{Castellan_Wezel_2013}. Fixing this exponent, we can fit our data to extract $T_c = 200.3$ K.

For the Cu$_{0.08}$TiSe$_2$ sample, correlated disorder creates a spatially varying local $T_c$. Cu poor patches of the sample should have a greater transition temperature than Cu rich patches. It is plausible that the underlying distribution of transition temperatures is Gaussian. In this case, the functional dependence of the intensity on temperature should be a convolution of the simple power law curve of a critical transition with this Gaussian distribution. This can be expressed as,
\begin{align}
I(T) \sim \int_{T}^{\infty} (T_c' - T)^{2\beta_c}\exp\bigg({-\frac{(T_m - T_c')^2}{2\Delta T^2}}\bigg) dT_c'
\label{eq:fit-curve}
\end{align}
\noindent where $T_m$ is the mean transition temperature and $\Delta T$ is the 1-$\sigma$ spread in transition temperatures. In this fit curve, we assume that the critical exponent of the power law curve inside the integral is the same as the pristine critical exponent: $2\beta_c$. In general, this assumption is not valid; however, changing the exponent simply shifts the best fit value of $T_m$ and alters $\Delta T$. If the exponent is left free in the least square fit, then it settles very close to zero. This then reduces the integral to a Gaussian convolution of a step function. Evidently, the functional form of Eq. \ref{eq:fit-curve} is capable of fitting the smeared transition for a wide range of input critical exponents.

For XPCS measurements it is important to quantify the correlation as indicated by the speckle contrast. A decay in this correlation indicates a dynamic system. The dynamics can be characterized by fitting the square of the intermediate scattering function to a compressed exponential:

\begin{equation}
F^2(\tau) = \exp\left(-\left(\frac{\tau}{\tau_F}\right)^\gamma\right)
\label{eq:fit-ce}
\end{equation}

\begin{figure*}
\includegraphics[width=1\linewidth]{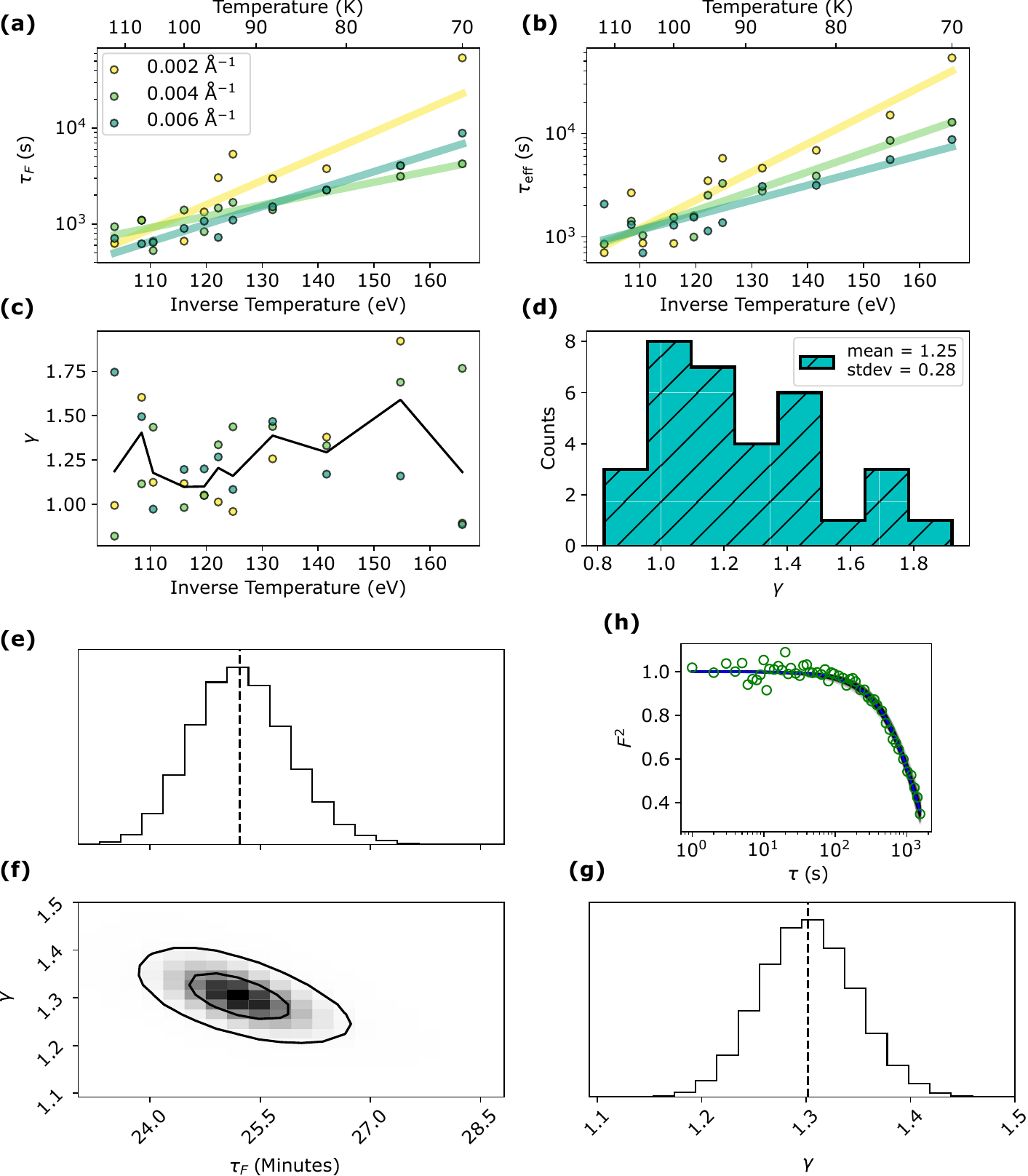}
    \caption{\textbf{Fit parameters of XPCS data on the ($\mathbf{0.5, \,0.5, \overline{4.5}}$) CDW peak of Cu$_{0.08}$TiSe$_2$.} The fits are to Eq. \ref{eq:fit-ce} with both $\tau_F$ and $\gamma$ as free parameters. (a) $\tau_F$ as a function of $1/(k_BT)$ for three different reciprocal space ROIs. The corresponding temperatures are shown on the upper axis. The outer radii of these tight packed annular ROIs are given in the legend. Solid lines are exponential fits (linear on log scale). (b) $\tau_{\text{eff}}$ defined in Eq.~\ref{eq:tau_eff}. (c) Fit values of $\gamma$. The solid, black line is an average of the three ROIs. (d) Histogram of all $\gamma$ values. (e)--(g) Corner plot of an MCMC fit of a stretched exponential to the correlation decay of the $0.004$ \AA$^{-1}$ ROI at 95 K temperature. The MCMC fit assumes a 0.025 1-$\sigma$ uncertainty in the measured values of $F^2$. Frames (e) and (g) show one dimensional histograms of $\tau_F$ and $\gamma$ respectively. Frame (f) shows a two dimensional histogram with the 1-$\sigma$ and 2-$\sigma$ confidence intervals plotted as solid curves. (h) Decay of $F^2$ along with a set of fit curves in gray corresponding to a random sample of the MCMC walkers. The blue curve is the best fit line.}
    \label{fig:all_fits}
\end{figure*}

Fig. \ref{fig:all_fits} shows fits to this equation with both $\tau_F$ and $\gamma$ as free variables. The data is analyzed in three close packed annular ROI regions (the lowest $q$ ROI is just a circle). The fits to $\gamma$ show no discernible temperature dependence and no $q$ dependence. If all of the data sets are included, then the mean value of $\gamma$ is 1.25 with a standard deviation of 0.28. Averaging the 3 ROIs gives a more precise measure of $\gamma$ for each temperature; this curve too indicates no obvious temperature dependence. The standard deviation of this ROI averaged measurement is 0.11. Fig. \ref{fig:all_fits}(d)--(g) illustrate a full MCMC fit of a compressed exponential to the 0.004 \AA$^{-1}$ ROI, 95 K dataset. Each data point in the fit is assumed to have a 0.025 1-$\sigma$ uncertainty in the value of $F^2$. The MCMC fit evolves 32 walkers through 5000 steps to construct a posterior distribution in $\tau_F$-$\gamma$ space. The results indicate that there is a negative cross-correlation between $\gamma$ and $\tau_F$ in the fit. This degeneracy results in a noisy plot of $\tau_F$ because the best fit value of $\gamma$ varies chaotically as a function of temperature and $q$. In the main text, the data are least squares fit to compressed exponential curves with $\gamma$ fixed at 1.2.

Another method to address this degeneracy problem is to compute an effective time scale variable ($\tau_{\text{eff}}$) that combines $\tau_F$ and $\gamma$. For a $\gamma = 1$ curve, the initial slope of the correlation decay is simply $-1/\tau_F$. However, this value is zero when $\gamma > 1$. Nevertheless, we can compute the average slope between $\tau=0$ and $\tau=f\tau_F$ where $f$ is some dimensionless number. This average slope then defines the effective timescale of the decay,
\begin{equation}
\tau_{\text{eff}}(f) = \frac{f \tau_F}{1-\exp(-f^\gamma)}
\label{eq:tau_eff}
\end{equation}
We choose $f=0.25$ for the plot of $\tau_{\text{eff}}$ in Fig. \ref{fig:all_fits}(b).

\begin{figure*}[h!]
\includegraphics[width=1.0\textwidth]{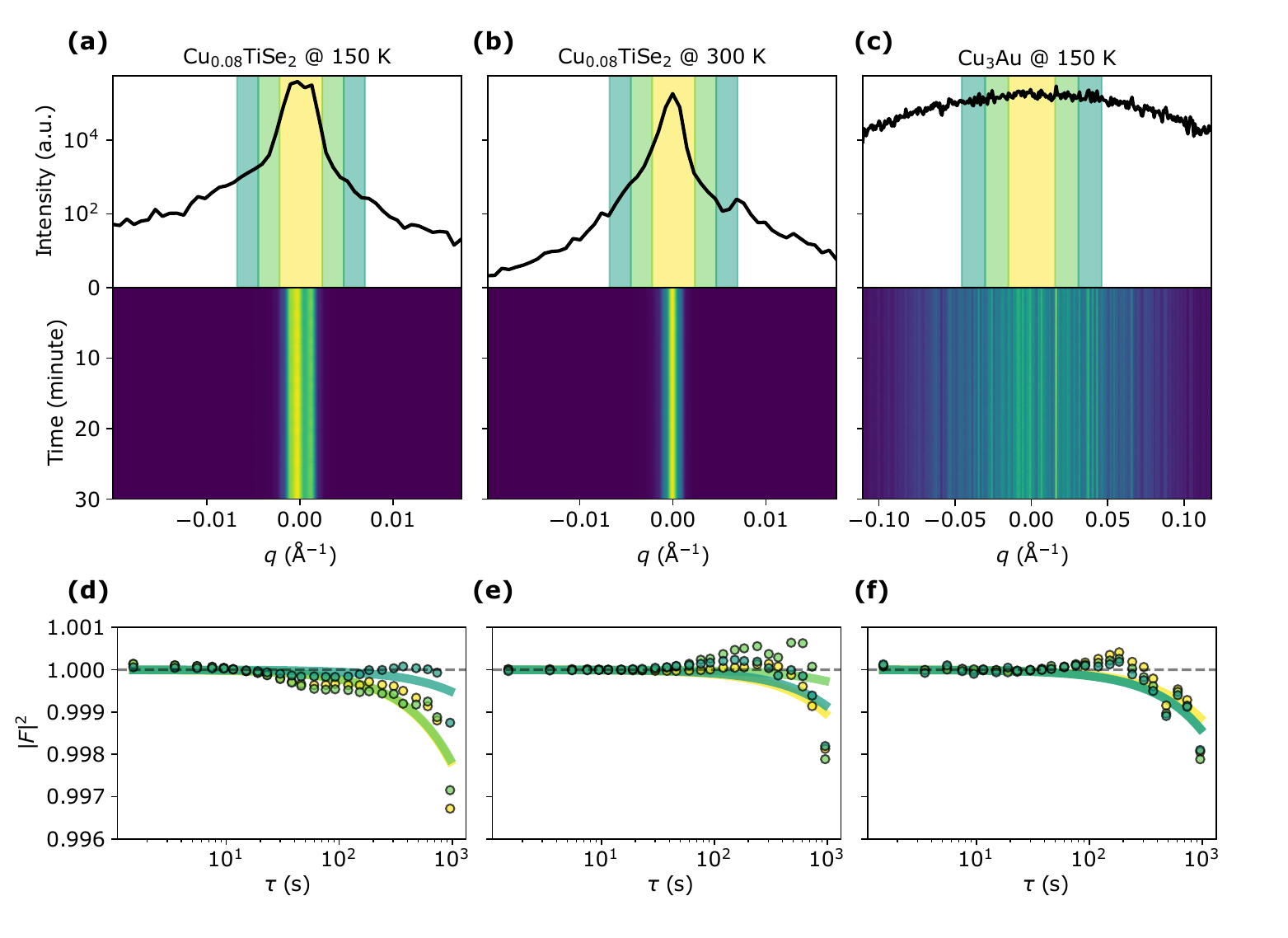}
    \caption{\textbf{XPCS on Bragg peaks} The 1st and 2nd columns present data from the (002) peak of Cu$_{0.08}$TiSe$_2$ at 150 K and 300 K respectively. The 3rd column is the (001) peak of Cu$_3$Au at 150 K. (a)--(c) Line cuts of the Bragg peaks. In the top panels, the shaded regions indicate different ROIs for XPCS, and the curves are average line cuts over all times. The bottom plots are waterfall images in which the line cut at a specific frame is rendered along the horizontal axis while the vertical axis is time. The duration of a single frame is 1.0 s. (d)--(f) The intermediate scattering function plotted against lag time. The colors of the curves indicate the associated ROIs shown in (a)--(c).}
    \label{fig:bragg-xpcs}
\end{figure*}

Supplementary Figure \ref{fig:bragg-xpcs} presents XPCS data on the (002) Bragg peak of Cu$_{0.08}$TiSe$_2$. The data exhibit a mostly static correlation function; the small decay in correlation that is observed can be fully attributed to the instrument by comparing to XPCS on a null sample. The crystal Cu$_3$Au serves as an excellent null sample for XPCS because it should exhibit no dynamics. Thus, an XPCS measurement on Cu$_3$Au measures the stability of the system and provides a maximum time window for reliability of the $g_2$ correlation function. In (c), XPCS on the 001 Bragg peak of Cu$_3$Au shows high stability over the full time window of $10^3$ s. Fits to exponential decays, give similar decay time constants for the 3 annular ROIs of $\tau_F \approx 7 \times 10^{5}$ s. The correlation decay for the Cu$_{0.08}$TiSe$_2$ Bragg peaks are comparable to this. If the Cu intercalates were mobile inside the Van der Waals gap, this would lead to a decay of the structural Bragg peak correlation. Thus, we can conclude that the Cu intercalates are static up to a temperature of 300 K.
%%TC:endignore

\section{Supplementary Section II. The Intermediate Scattering Function}

The normalized intermediate scattering function is related to the one-time correlation function according to the equation: $g_2(\vec{q}, \tau) = 1 + \beta \vert F(\vec{q}, \tau)\vert^2$ where $\beta$ is the speckle contrast factor. This quantity is associated with the field correlation function $g_1(\vec{q}, \tau)$ through the Siegert relation \cite{dls}. For the system considered here (a triple CDW), the intermediate scattering function can be expressed directly in terms of the CDW order parameter. Let the charge density be denoted as $\rho(\vec{x})$. Then, we may define an order parameter $\alpha(\vec{x})$ as the normalized charge density wave atop the original lattice charge distribution,
\begin{align}
\rho(\vec{x}) = \rho_0(\vec{x})(1 + \alpha(\vec{x}))
\end{align}
This component combines the three separate CDWs as,
\begin{align}
\alpha(\vec{x}, t) = \text{Re}\left(\psi_1(\vec{x}, t) + \psi_2(\vec{x}, t) + \psi_3(\vec{x}, t)\right)
\end{align}
The intermediate scattering function is then related to the spatial Fourier transform of this order parameter according to,
\begin{align}
F(\vec{k}, \tau) \propto \langle \tilde{\alpha}(\vec{k}, t + \tau) \tilde{\alpha}(-\vec{k}, t) \rangle_t
\end{align}
where the angled brackets indicate an average over $t$ at equilibrium (i.e., an ensemble average for an ergodic system).

For simplicity, we consider a single CDW modulation where the order parameter can be represented as $\alpha(\vec{x}, t) = A(\vec{x}, t) \cos({\vec{k}\,}'\cdot\vec{x} + \phi(\vec{x}, t))$. In the phase fluctuation state of $T_c$ disorder, much of the bulk has $A = 0$. The ordered regions have roughly constant $A$ and uniform (but dynamic) phase. In this case, we can express $\tilde{\alpha}$ as a sum over the ordered regions ($D_j$),
\begin{align}
\tilde{\alpha}(\vec{k}, t) = A\sum_{j}\int_{D_j} d^3x~e^{-i\vec{k}\cdot\vec{x}}\cos({\vec{k}\,}'\cdot\vec{x} + \phi_j(t))
\end{align}
This expression yields peaks at $\vec{k} = \pm {\vec{k}\,}'$; however, in a scattering experiment, we only consider one of these peaks. So, we can replace the cosine term with just the $+{\vec{k}\,}'$ complex exponential and define the expression in terms of the momentum transfer: $\vec{q} = \vec{k} - {\vec{k}\,}'$. The expression then simplifies to,
\begin{align}
\tilde{\alpha}(\pm\vec{k}, t) \approx \frac{A}{2}\sum_{j}e^{\pm i\phi_j(t)} \int_{D_j} d^3x~e^{\mp i\vec{q}\cdot\vec{x}}
\end{align}

Given that the $T_c$ disorder is quenched, it is assumed that $D_j$ is a static region. Thus, the integral can be reduced to a function of $\vec{q}$,
\begin{align}
w_j(\vec{q}\,) = \int_{D_j} d^3x~e^{- i\vec{q}\cdot\vec{x}}
\end{align}
Finally, the intermediate scattering function (evaluated near $\vec{k} = {\vec{k}\,}'$) can be expressed as a function of $\vec{q}$,
\begin{align}
F(\vec{q}, \tau) \propto  \sum_{j,j'} w_j(\vec{q}\,)w_{j'}(-\vec{q}\,)  \left\langle \exp\bigg(i(\phi_j(t + \tau) - \phi_{j'}(t))\bigg) \right\rangle_t
\end{align}

The correlation function between the phase terms should only depend on the index $j$ because the $j'$ term has no $\tau$ dependence. In the case of a commensurate, unit-cell doubling CDW, the phase $\phi$ is either $0$ or $\pi$. The telegraph noise type dynamics expected for a $T_c$ disordered system results in a compressed exponential decay of the ensemble correlation function (see the discussion of a random telegraph noise model in Supplementary Section IV). 
\begin{align}
c_{j}(\tau) &= \left\langle \exp\bigg(i(\phi_j(t + \tau) - \phi_{j'}(t))\bigg)\right\rangle_t \\
&= \exp\bigg(-(\Gamma_j\tau)^{\gamma_j}\bigg)
\end{align}
Here, $\Gamma_j$ is the flipping rate, and the exponent $\gamma_j$ is related to the ballistics of the phase flip process (see Supplementary Figure \ref{fig:telegraph-model}). The intermediate scattering function can now be expressed as,
\begin{align}
F(\vec{q}, \tau) &\propto  \Omega(-\vec{q}\,)\sum_{j} w_j(\vec{q}\,)  c_j(\tau) \\
\Omega(\vec{q}\,) &= \sum_j w_j(\vec{q}\,)
\end{align}
Clearly, the weight term $w_j(\vec{q}\,)$ is a Dirac-delta function centered at $\vec{q} = 0$ when the ordered region $D_j$ is of infinite extent. For a finite sized $D_j$, the function $w_j(\vec{q}\,)$ broadens.

This can be solved exactly in three-dimensions for spherical $D_j$. Let $L_j$ be the radius of $D_j$. Then, the weight is a function of $q=\lvert\vec{q}\,\rvert$,
\begin{align}
w_j(q) = 4\pi \frac{\sin(qL_j)-qL_j\cos(qL_j)}{q^3}
\end{align}
The function has a removable singularity at $q=0$ which can be evaluated by limit: $\lim_{q\rightarrow 0}(w_j(q)) = \frac{4\pi}{3}L_j^3$. So, the contribution at $q=0$ is the volume of the ordered region. The width of the function can be quantified by finding the standard deviation ($\Delta q_j$) of the Gaussian that has a curvature at $q=0$ equal to that of $w_j(q)$. This is given by $\Delta q_j = \sqrt{5}/L_j$.
Thus, larger ordered regions contribute disproportionally to $F(\vec{q}, \tau)$ at low $q$, and smaller ordered regions dominate the signal at high $q$. The flipping rate for an ordered region $\Gamma_j$ is a monotonically decreasing function of $L_j$; thus, $F(\vec{q}, \tau)$ develops a $q$ dependence where the overall decay time parameter $\tau_F$ monotonically decreases with $q$.

\section{Supplementary Section III. Correlated Disorder Ising Model Simulation}

A standard 3D Ising model simulation consists of a grid of sites on a cubic lattice with cyclic boundary conditions. A given site $(i, j, k)$ can exist in one of two possible states: $\psi_{ijk} = \pm 1$. Each site is linearly coupled to its six nearest neighbors with some coupling constant $J$. The energy of the interaction between nearest neighbors $(i, j, k)$ and $(i', j', k')$ is conventionally given by $-2 J \psi_{ijk}\psi_{i'j'k'}$. Thus, each site can be assigned an energy based on a sum of the states of its nearest neighbors defined as,
\begin{align}
\sigma_{ijk} &= \psi_{(i+1)jk} + \psi_{(i-1)jk} + \psi_{i(j+1)k} + \psi_{i(j-1)k} + \psi_{ij(k+1)} + \psi_{ij(k-1)}. \\
E_{ijk} &= -2 J \psi_{ijk}\sigma_{ijk}.
\end{align}

This energy is then used to compute a transition probability $\Pi_{ijk}$ for site $(i, j, k)$. Here, we use a sigmoid function of temperature $T$ as the transition probability defined as,
\begin{align}
\Pi_{ijk} = \frac{1}{1 + \exp\left(\frac{2}{T}\sigma_{ijk} \psi_{ijk} J\right)}.
\label{eq:trans-1}
\end{align}

In the typical Metropolis--Hastings algorithm, the transition probability is 100$\%$ for steps that reduce energy, and steps that increase energy have a probability determined by a Boltzmann factor. The sigmoid function simulation method used here is known as the Glauber algorithm; it has a slower convergence but is more appropriate to a simulation that aims to simulate dynamics \cite{Glauber_1963}.

The simulation proceeds by a series of discrete steps. In each step, a site is randomly chosen and its state is flipped from $\pm 1$ to $\mp 1$ with probability $\Pi_{ijk}$; otherwise, the state is left unchanged. The system is initialized with each site randomly assigned to either $+1$ or $-1$. In this pristine 3D Ising Model with a coupling constant of $J = 1$, the system becomes ordered below a simulation temperature of $T \approx 4.52$ (in 2D it is $T \approx 2.27$). Exactly at the transition temperature, the system is in a fractal state in which fluctuations occur at all length scales and time scales. This model has been thoroughly studied with numerical techniques and was solved analytically in 2D by Onsager \cite{Onsager_1944}. Here, we explore an extension of the Ising model to incorporate correlated disorder.

In general, one can introduce two forms of disorder into the transition probability expression of Eq. \ref{eq:trans-1}: a spatially varying field $h_{ijk}$ that linearly couples to the state $\psi_{ijk}$, and a variable site coupling $J_{ijk}$. This results in a more general expression for the transition probability given by,
\begin{align}
\Pi_{ijk} = \frac{1}{1 + \exp\left(\frac{2}{T}\left(\sigma_{ijk} + h_{ijk}\right)\psi_{ijk} J_{ijk}\right)}.
\end{align}

\begin{figure*}
\includegraphics[width=1.0\textwidth]{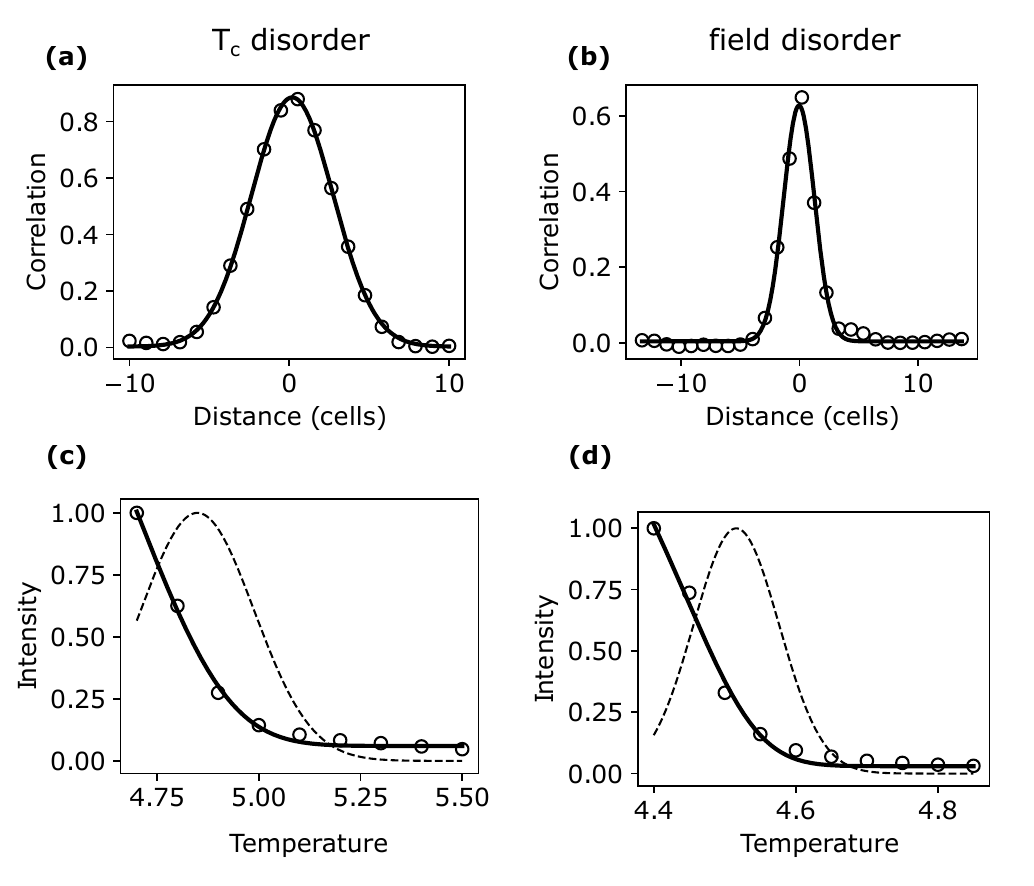}
    \caption{\textbf{Properties of the 3D \textit{T}$\mathbf{_c}$ disorder and field disorder simulations.} (a) Spatial correlation of the coupling constant for the $T_c$ disorder model as given by equation \ref{eq:corr}. The 1-$\sigma$ correlation is 2.56 cells. (b) Spatial correlation of the disorder for the field disorder model. The 1-$\sigma$ correlation is 1.30 cells. (c)--(d) Mean-square model magnetization against temperature for the $T_c$ disorder and field disorder simulations respectively. The solid lines are fits to equation \ref{eq:fit-curve} while the dashed line is a Gaussian illustrating the parameters of the fit. For (b), the fit values are $T_m = 4.85$ and $\Delta T = 0.14$. For (c), the fit values are $T_m = 4.52$ and $\Delta T = 0.06$.
    }
    \label{fig:3d-model-summary}
\end{figure*}

The term $h_{ijk}$ introduces field disorder while $J_{ijk}$ is associated with $T_c$ disorder. The correlation of the disorder can be quantified by a spatial correlation function of either $h_{ijk}$ or $J_{ijk}$. This is expressed in terms of $J$ as,
\begin{align}
C_{ijk} = \frac{\langle J_{i'j'k'}J_{(i'+i)(j'+j)(k'+k)} \rangle}{\langle J_{i'j'k'}^2  \rangle}
\end{align}
\noindent where the angled bracket denote an average over $i'$, $j'$, and $k'$. For the simulations discussed in this work, the correlation function is approximately a Gaussian in the variable $r^2 = i^2 + j^2 + k^2$. This is expressed as,
\begin{align}
C(r) = \exp\bigg({-\frac{r^2}{2\Delta^2}}\bigg) \label{eq:corr}
\end{align}
\noindent where $\Delta$ is the correlation length of the disorder. Line cuts of the disorder correlation functions for the $T_c$ and field disorder simulations are given in Figure \ref{fig:3d-model-summary}(a)--(b). The field disorder correlation length is smaller than the $T_c$ disorder correlation length because the field disorder has a random sign while the temperature disorder is strictly positive. In Supplementary Fig.~\ref{fig:3d-model-summary}(c)--(d), the mean-square of the magnetization is plotted against temperature for the two models and fit to equation \ref{eq:fit-curve}. The fit values for the $T_c$ disorder model are $T_m = 4.85$ and $\Delta T = 0.14$. The fit values for the field disorder model are $T_m = 4.52$ and $\Delta T = 0.06$.

For the $T_c$ disorder model, correlated disorder in $J$ will result in various regions of the sample undergoing the phase transition at different temperatures. A patch of the sample for which the average coupling constant is $J'$ will have a local transition temperature of $T_c' = J'T_c$. The simulations presented in this work have $J_{ijk}$ distributions in which the background value is 1 and the disorder is strictly of $J_{ijk} > 1$. Thus, the disorder in the simulations always increases the local transition temperature. At intermediate temperature, islands of order will form in the regions of high coupling constant while the bulk of the system remains disordered. These islands will be subject to perturbations at their boundaries due to the thermal fluctuations of the surrounding disorder. These thermal fluctuations can flip the state of some fraction of the ordered region resulting in ballistic propagation of a domain wall that will fully convert that region. This type of thermally activated dynamics occurs spontaneously in equilibrium at these intermediate temperatures.

In order to produce simulated XPCS data from this Correlated Disorder Ising Model it is necessary to generate a ``diffraction pattern'' from the Ising model state. This is readily accomplished by taking the modulus squared of the multidimensional Fast Fourier Transform (FFT) of the system state $\psi_{ijk}$. In real data, a diffraction pattern is collected over some exposure time ($1.0$ s in this work) in which fast variations in the speckle contrast are averaged away. In simulation, a diffraction image is given an exposure time as some number of simulation steps $N$. The $n$\textsuperscript{th} saved diffraction image is then,
\begin{align}
\text{I}_n = \sum_{s = 1 + nN}^{(n+1)N}\bigg\lvert\text{FFT} \left(\psi^{(s)}\right)\bigg\rvert^2
\end{align}
\noindent where $\psi^{(s)}$ is the system state at the $s$\textsuperscript{th} simulation step. In practice, it is more computationally efficient to only compute an FFT of the state every $M$ many steps where $M = L^3$ and $L$ is the side length of the simulation grid. This greatly speeds up computation without appreciably changing the results. On the machine used for the simulations in this study, the computation speed is approximately $0.1$ $\mu$s/step where a single step is one binary decision on whether to flip a randomly chosen spin.

\new{The 3D model discussed in the main text reproduces the thermally activated dynamics observed in the experimental data. Extraction of $\tau_F$ values using Eq.~{\ref{eq:fit-ce}} yields $\tau_F = \tau_0 \exp(\Delta E / T)$, where $T$ is the simulation temperature. To make the fitted $\Delta E$ values from simulation directly comparable to experiment, we express them in units of the simulation's $T_c$, and compare them to the experimental values, which are given in units of $k_B T^{\text{Cu}}_{\text{CDW}}$. As shown in Supplementary Table~{\ref{tab:data_vs_sim}}, the simulation reproduces the same trend in $q$ for both $\Delta E$ and $\tau_0$, and yields comparable $\Delta E$ values when expressed in units of $T_c$.}

\begin{table}[h]
\centering
\caption{Comparison of Arrhenius fit parameters from experiment and simulation.}

% Adjust spacing
\renewcommand{\arraystretch}{1.3} % vertical space
\setlength{\tabcolsep}{12pt}      % horizontal space

\begin{tabular}{c|ccc|ccc}
\hline
 & \multicolumn{3}{c|}{Data} & \multicolumn{3}{c}{Simulation} \\
ROI & $q$ (\AA$^{-1}$) & $\Delta E$ ($k_BT^{\text{Cu}}_{\text{CDW}}$) & $\tau_0$ (s) & $q$ (px) & $\Delta E$ ($T_c$) & $\tau_0$ (steps) \\
\hline
1 & 0.002 & 8.62 & 2.03  & 1 & 8.41 & $7.07 \times 10^{-9}$ \\
2 & 0.004 & 6.58 & 8.37  & 2 & 7.77 & $8.52 \times 10^{-8}$ \\
3 & 0.006 & 5.33 & 17.90 & 3 & 6.59 & $1.18 \times 10^{-5}$ \\
\hline
\end{tabular}
\label{tab:data_vs_sim}
\end{table}

\begin{figure*}
\includegraphics[width=1.0\textwidth]{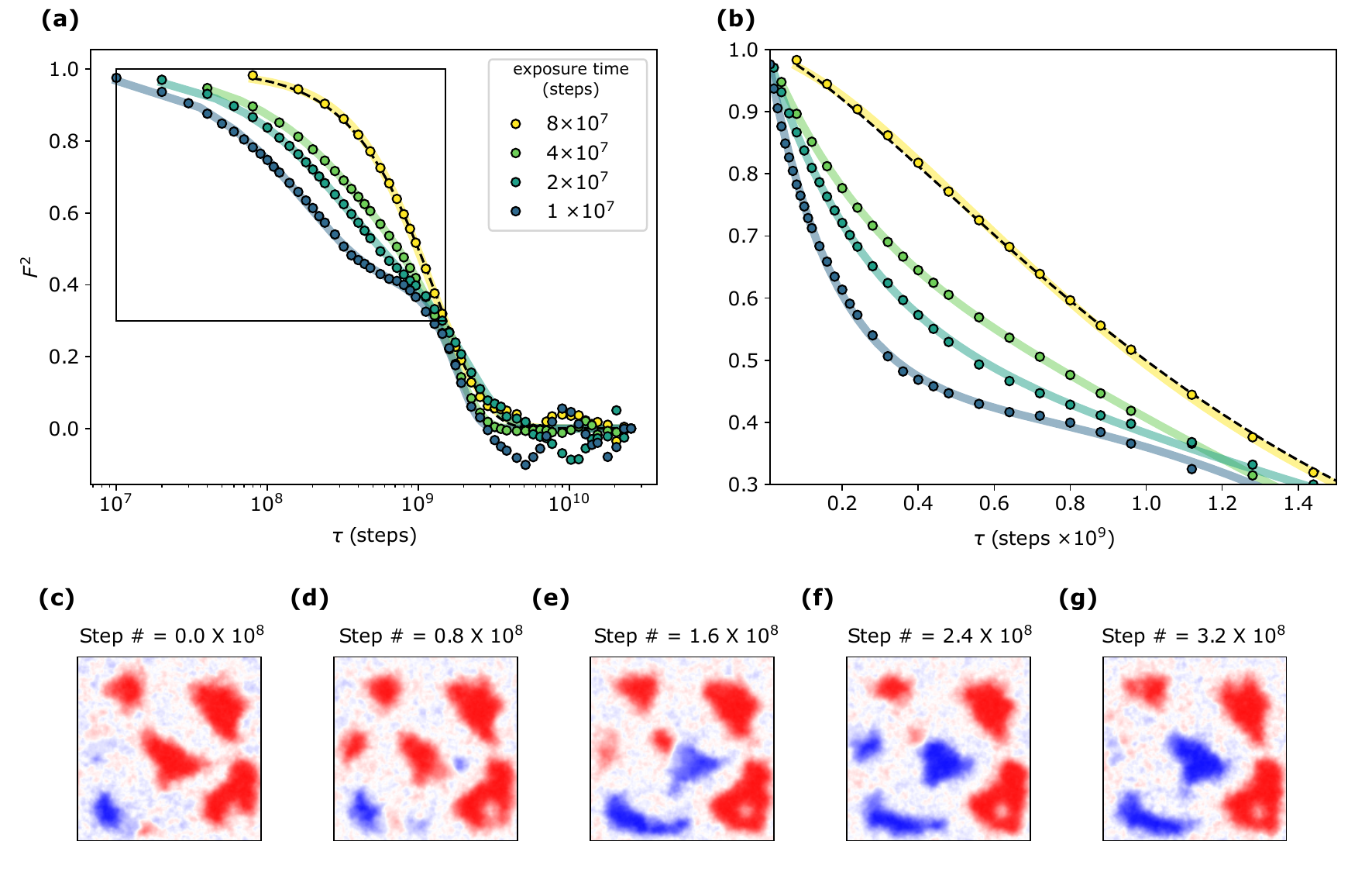}
    \caption{\textbf{Quasi-ballistic dynamics in a correlated disorder Ising model.} The results presented here are for a simulation on a 200 $\times$ 200 grid at a simulation temperature of 2.8. with a peak $J_{ij}$ of 1.4 and a disorder correlation length of $\Delta = 12.6$. (a) The intermediate scattering function computed for different simulation exposure times as indicated in the legend. The dashed black line is a compressed exponential fit to the highest exposure of $8 \times 10^7$ with $\gamma = 1.32$. The thick, solid lines are fits to the sum of a stretched exponential and a compressed exponential. (b) Linear scale plot of the region indicated by a black square in (a). (c)--(g) Real space snapshots of the simulation state at the indicated step numbers. The frames show the ballistic conversion of one of the ordered regions.}
    \label{fig:exposures}
\end{figure*}

To construct a model with a correlation decay that is clearly of compressed rather than simple exponential form, it is necessary to simulate on a large grid size. Consider Supplementary Fig.~\ref{fig:exposures} which presents results from a 2D simulation on a 200 $\times$ 200 grid. In this case, low exposure times show two separate timescales in the correlation function decay. The square of the intermediate scattering function computed at the peak center fits to,
\begin{align}
F^2 = A\exp\left(-\left(\frac{t}{\tau_1}\right)^{\gamma_1}\right) + (1-A)\exp\left(-\left(\frac{t}{\tau_2}\right)^{\gamma_2}\right)
\end{align}
\noindent where $\gamma_1 < 1$ and $\gamma_2 > 2$. The stretched exponential component of the decay comes from fast fluctuations throughout the system and fluctuations of the ordered region boundaries. The comparatively slower compressed exponential component comes from the flipping of the ordered regions. As the exposure time is increased, the stretched exponential component is suppressed until the curve converges to a single compressed exponential with $\gamma = 1.32$. The source of this super-diffusive dynamics is the ballistic domain wall motion involved in the conversion process of the ordered regions. This is illustrated in the progression of frames shown in Supplementary Fig.~\ref{fig:exposures}(c)--(g).

%%%

Correlated $T_c$ disorder can also promote the formation of metastable domain walls. Consider the simulations presented in Supplementary Fig.~\ref{fig:domain_wall}. This compares the three dimensional pristine and $T_c$ disorder simulations on a larger 100 $\times$ 100 $\times$ 100 cell grid. The simulations are initially equilibrated for $10^9$ steps at a simulation temperature of $T = 4.7$ (dimensionless units); for the pristine system, $T_c \approx 4.51$. After this initial equilibration, the pristine and $T_c$ disorder simulation temperatures are reduced at a constant rate of $\Delta T/T_c = -2 \times 10^{-12}$ per step ($T_c$ here being the pristine system value). The simulations were run until reaching a final temperature of $T/T_c = 0.962$.

The pristine system quickly establishes long range order once the temperature is below $T_c$. However, the $T_c$ disordered system shows a more complex behavior. As the system approaches the global $T_c$, bridges form between the ordered regions. These then expand until regions of opposite phase meet at domain walls. This domain wall formation mechanism is robust even with the very slow cooling rate of this simulation. The domains formed through this process are metastable because the true ground state of the system is still a single ordered region. Nevertheless, in a physical system, these domains can be long lived if the system loses ergodicity (i.e., the time average of the system is no longer a thermodynamic ensemble average).

\section{Supplementary Section IV. Random Telegraph Noise Model}

A simple model of telegraph noise can be used to explore the physical meaning of the exponent $\gamma$ in the compressed exponential. In the full Ising model simulations, a sum of $\psi_{ijk}$ over an ROI inside an ordered region produces a telegraph noise signal. Here, we model this ROI-averaged value with a simpler model of a single scalar quantity $\Psi (t) \in [-1,+1]$. The variable $t$ is discrete time as measured in number of steps. This quantity is initialized randomly at a value of either $+1$ or $-1$. It has a constant probability per unit time of initiating a sign flip: $\Gamma = 1/t_{\text{flip}}$. When a flip is initiated, it takes some duration $t_\text{conv}$ to complete. During this interval, $\Psi (t)$ changes linearly to its new value. For instance, suppose $\Psi(t_0) = -1$ and a flip is triggered at $t_0$. Then $\Psi (t) = -1 + 2(t-t_0)/t_{\text{conv}}$ for $t_0 < t < t_0 + t_{\text{conv}}$. For $t \geq t_0 + t_{\text{conv}}$, the value of $\Psi(t)$ is stable at $+1$ until another flip is initiated. The analogous process occurs for a flip from $+1$ to $-1$ in which the state value linearly decreases to its new stable point over a duration of $t_{\text{conv}}$. During the conversion process, the probability per step of flipping to a different stable point is still $\Gamma$; in this case, the system will turn mid transition and begin moving linearly back to its previous stable point.

The results of the simulation can be characterized by computing a temporal auto-correlation function on $\Psi(t)$ defined as,
\begin{align}
g_1(\tau) = \frac{\langle \Psi(t)\Psi(t + \tau) \rangle}{\langle \Psi(t)^2 \rangle}
\end{align}
\noindent where the angled brackets denote an average over $t$. This quantity can be associated with the intermediate scattering function. In the context of this simulation, time is discrete; so, the minimum conversion time is 1 step. In this limit, the correlation decay can be derived analytically: $g_1(\tau) = \exp(-(\Gamma \tau/2)^{\gamma})$ with $\gamma = 1$. This is purely diffusive dynamics. However, when we increase $t_{\text{conv}}$, the dynamics become quasi-ballistic because the system spends some time in linear motion between set points. The correlation decay then fits to a compressed exponential curve ($\gamma > 1$). This is demonstrated in Supplementary Fig.~\ref{fig:telegraph-model}(a) in which results from a collection of simulations at various values of $\Gamma$ and $t_{\text{conv}}$ are presented. The fit parameter $\gamma$ monotonically increases as a function of the dimensionless quantity $\Gamma t_{\text{conv}}$ between 0 and 1. At $\Gamma t_{\text{conv}} = 1$, the model is highly ballistic; the state is almost always in linear motion between the stable points and no longer resembles telegraph noise. In this work, the XPCS data yielded an average value of $\gamma = 1.23$. For this model, such a value is obtained for $\Gamma t_{\text{conv}} = 0.24$. If the conversion of ordered regions proceeds from a moving front, then the average CDW phase will change linearly between the two possible values (i.e., 0 and $\pi$). This is a likely mechanism for the observed quasi-ballistic dynamics.

\bibliography{refs_sup.bib}
\newpage{}

\section{Supplementary Figures}

\begin{figure*}[h!]
\includegraphics[width=1.0\textwidth]{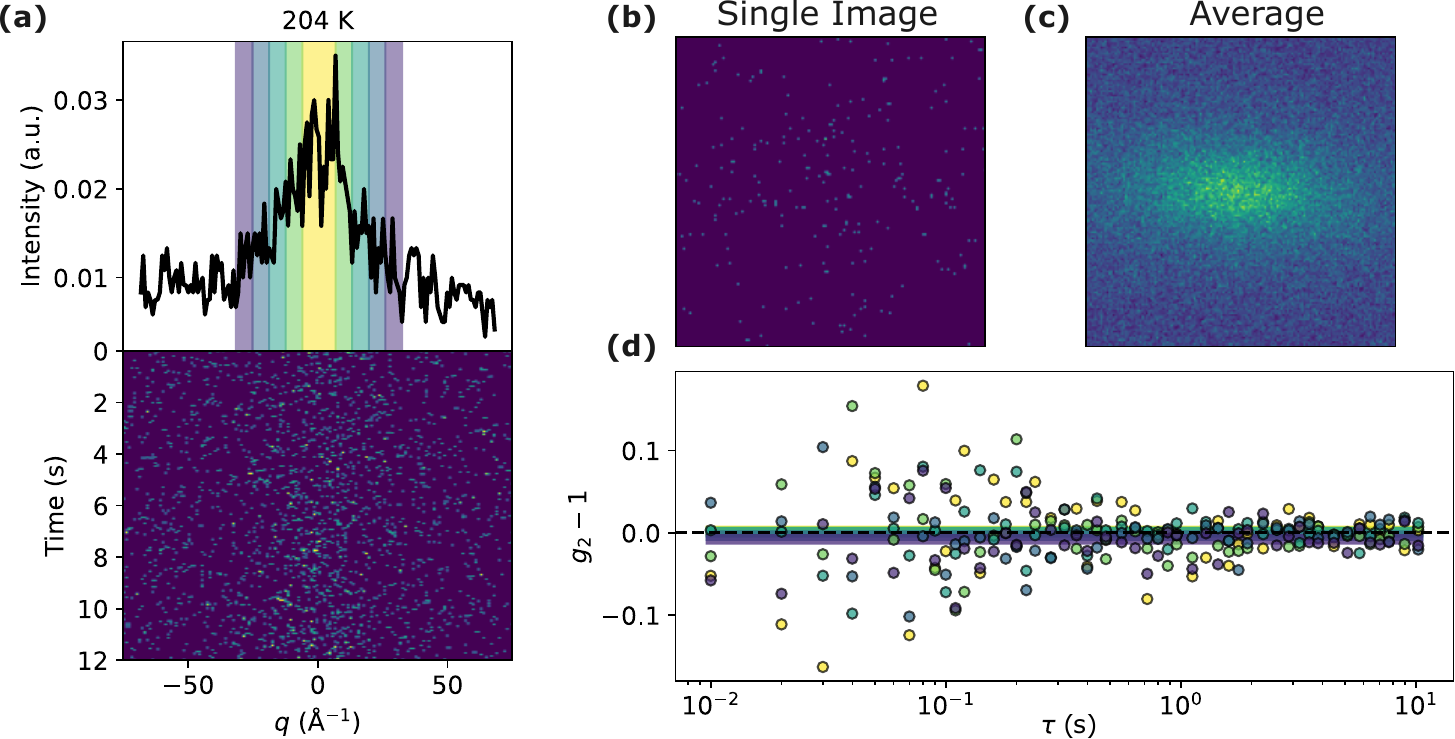}
    \caption{\textbf{XPCS with 10 ms exposure time on a CDW peak of pristine 1\textit{T}-TiSe$\mathbf{_2}$}. All data is in reference to the (0.5, 0.5, $\overline{4.5}$) peak at a temperature of 204 K. (a) Line cut along $h$ of the CDW peak. The shaded regions are different ROIs for XPCS. The bottom panel is a linear scale waterfall image in which the line cut at a specific frame is rendered along the horizontal axis while the vertical axis is time. The duration of a single frame is 10 ms. (b) Single frame image of the CDW peak. (c) Average of 1200 images. (d) Speckle correlation as a function of lag time. The colors of the curves indicate the associated ROIs shown in (a)}    \label{fig:10_ms_data}
\end{figure*}

\begin{figure*}[h!]
\includegraphics[width=1.0\textwidth]{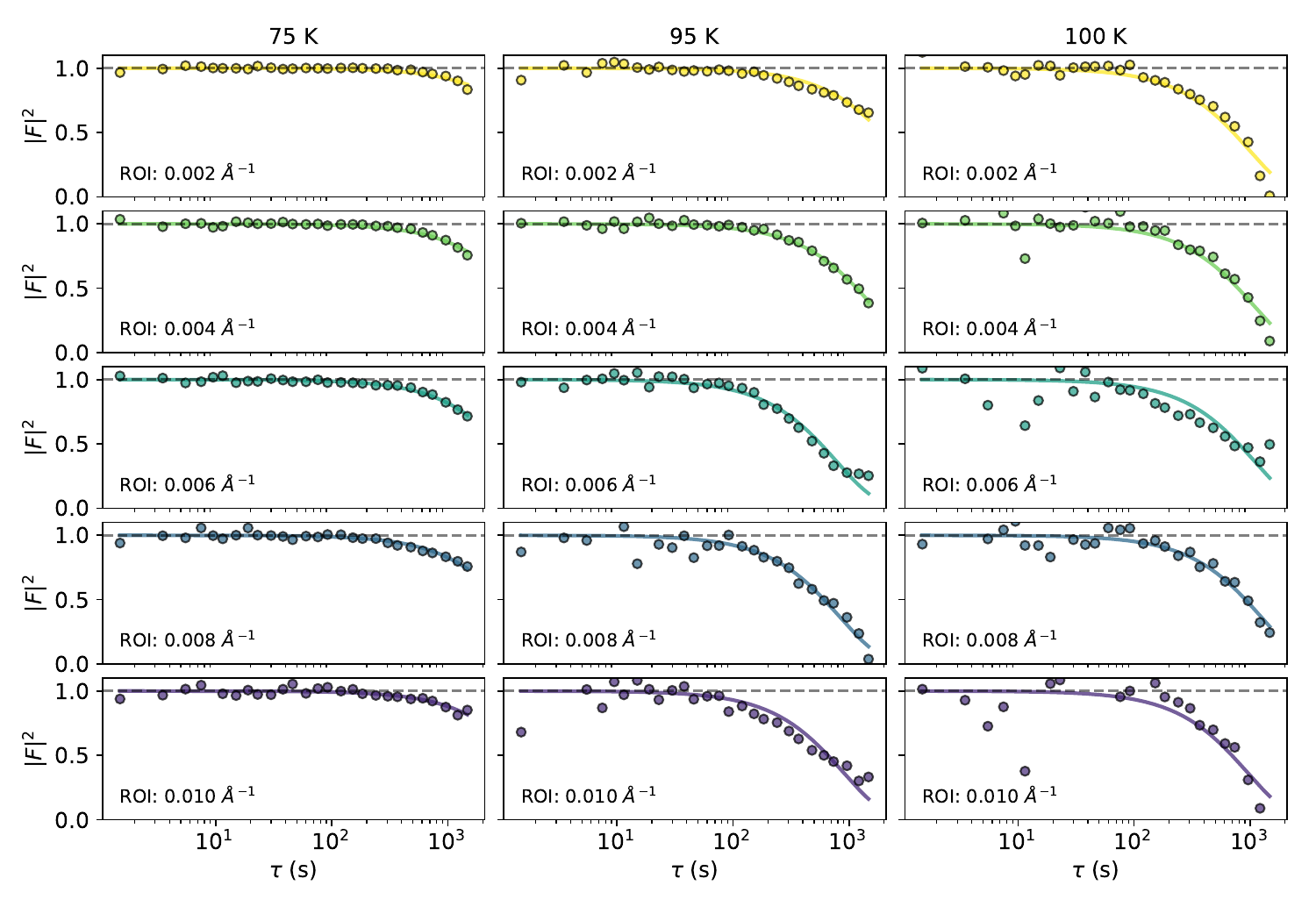}
    \caption{\textbf{XPCS on a CDW peak of Cu$_{\mathbf{0.08}}$TiSe$\mathbf{_2}$ (separated ROIs)}. Each panel shows the intermediate scattering function plotted against lag time. All data is in reference to the ($0.5$, $0.5$, $\overline{4.5}$) CDW peak. The 1st, 2nd, and 3rd columns show data at 75 K, 95 K, and 100 K respectively. The five different rows correspond to different reciprocal space ROIs; the outer radii of these ROIs are indicated in the panels. Trend lines are fits to a compressed exponential with fixed $\gamma$ = 1.2.}
    \label{fig:F2_all_ROIS}
\end{figure*}

\begin{figure*}[h!]
\centering
\includegraphics[width=0.7\textwidth]{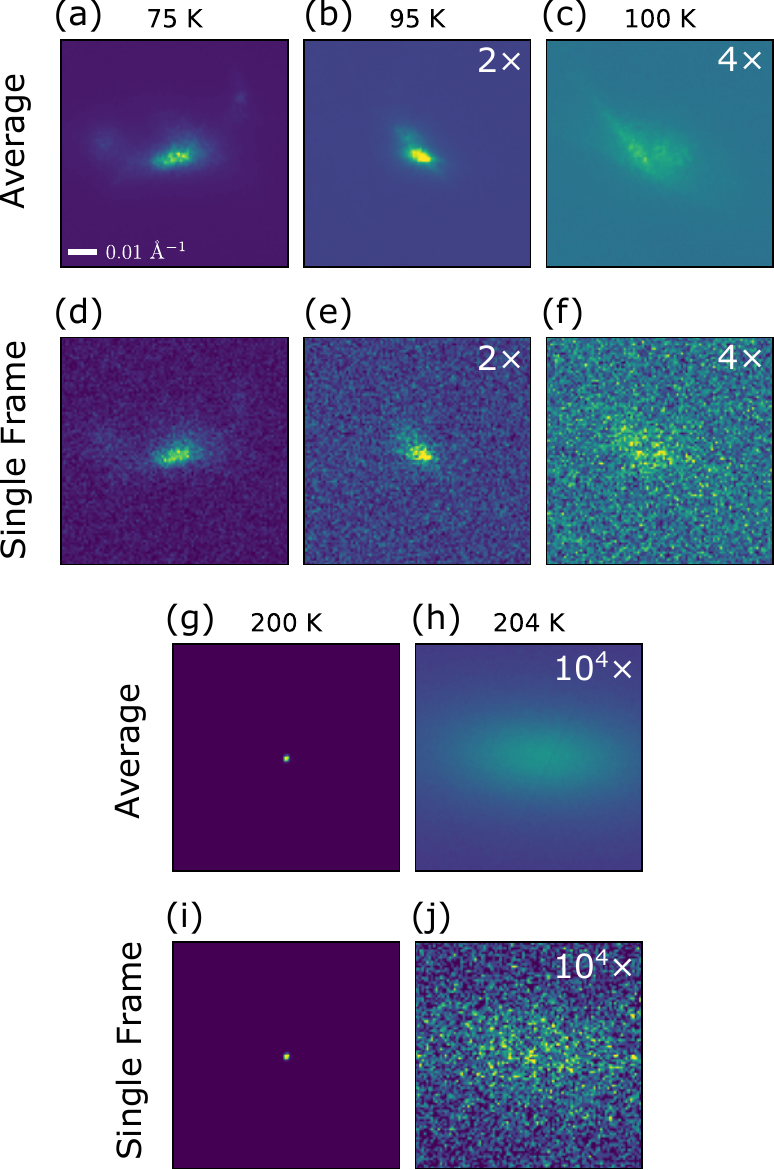}
    \caption{\textbf{XPCS diffraction patterns at the ($\mathbf{0.5, \,0.5, \overline{4.5}}$) CDW peak.} 
    (a)--(c) Averaged diffraction patterns from 1800 images (1.0 s exposure each) at 75 K, 95 K, and 100 K, respectively, for the Cu$_{0.08}$TiSe$_2$ sample. 
    (d)--(f) Single-frame diffraction patterns corresponding to (a)--(c). 
    (g)--(h) Averaged diffraction patterns (1800 images) for the pristine system at 200 K and 204 K, respectively. 
    (i)--(j) Single frames corresponding to (g)--(h).}
    \label{fig:speckle_pattern_images}
\end{figure*}

\begin{figure*}
\includegraphics[width=1.0\textwidth]{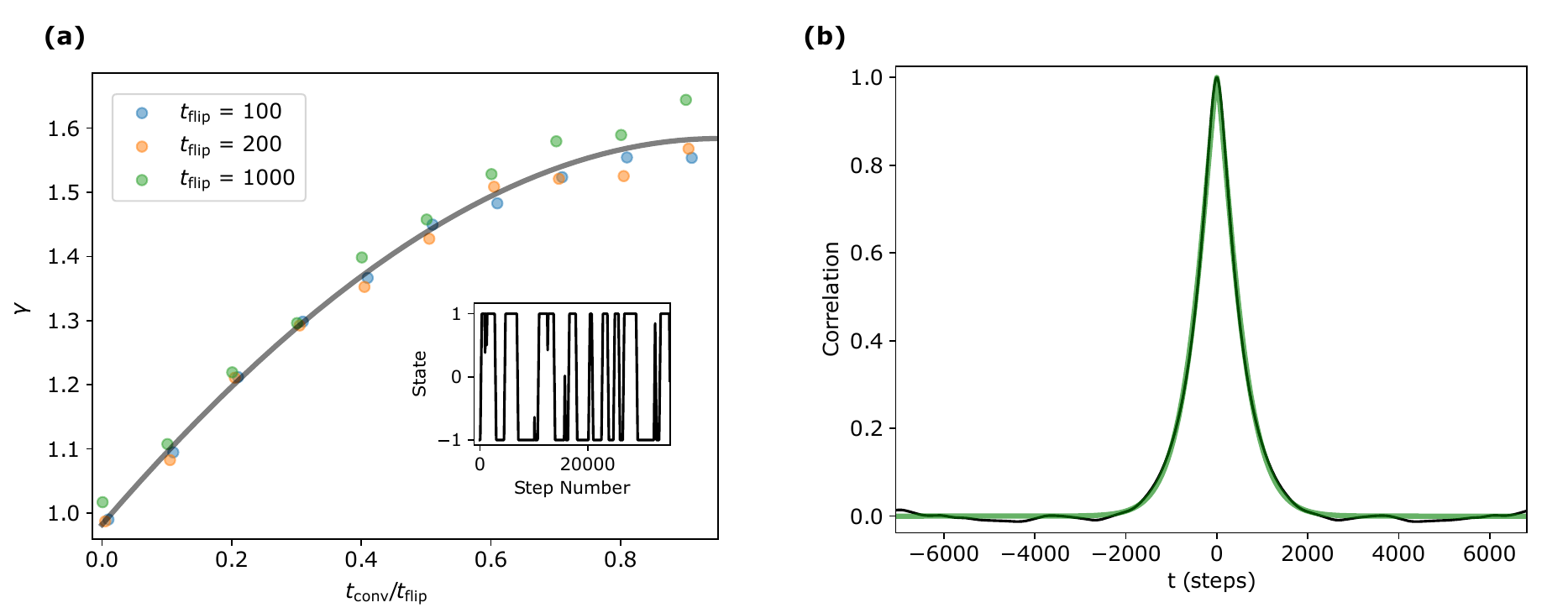}
    \caption{\textbf{Random telegraph noise model} (a) Fit value of the exponent in a compressed exponential as a function of the dimensionless conversion time. The three separate scatter plots are for different values of the flipping rate ($1/\Gamma = t_{\text{flip}}$) as indicated in the legend. The solid, gray line is a 3rd degree polynomial fit to the average of the 3 scatter plots. The inset shows an example section of the binary state evolution. (b) State autocorrelation function with fit to a compressed exponential decay having $\gamma = 1.2$. The inset plot in (a) and the plot in (b) are for the same simulation with $t_{\text{flip}} = 1000$ and $t_{\text{conv}} = 250$.}
    \label{fig:telegraph-model}
\end{figure*}

\begin{figure*}
\includegraphics[width=1.0\textwidth]{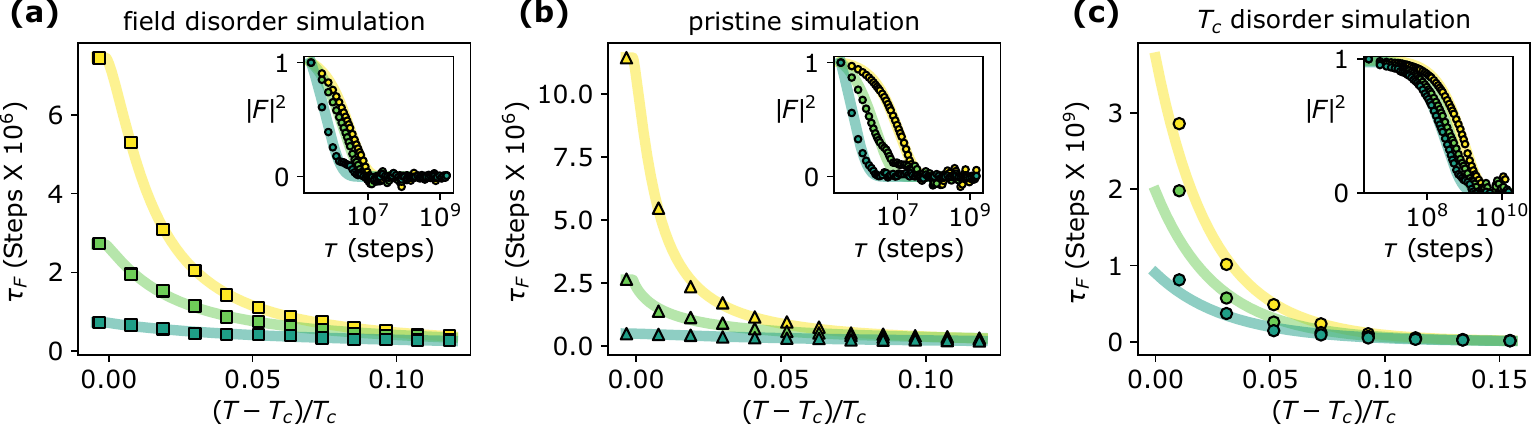}
    \caption{\textbf{Simulated field disorder, pristine, and \textit{T}$\mathbf{_c}$ disorder comparison}. The panels show the fit values of $\tau_F$ as a function of temperature. (a)--(c) show results from the field disorder, pristine, and $T_c$ disorder simulations respectively. The insets show the intermediate scattering function against step number at the nominal $T=T_c$ simulation temperature.}
    \label{fig:field_Tc_pristine_compare}
\end{figure*}

\begin{figure*}[t]
\includegraphics[width=1.0\textwidth]{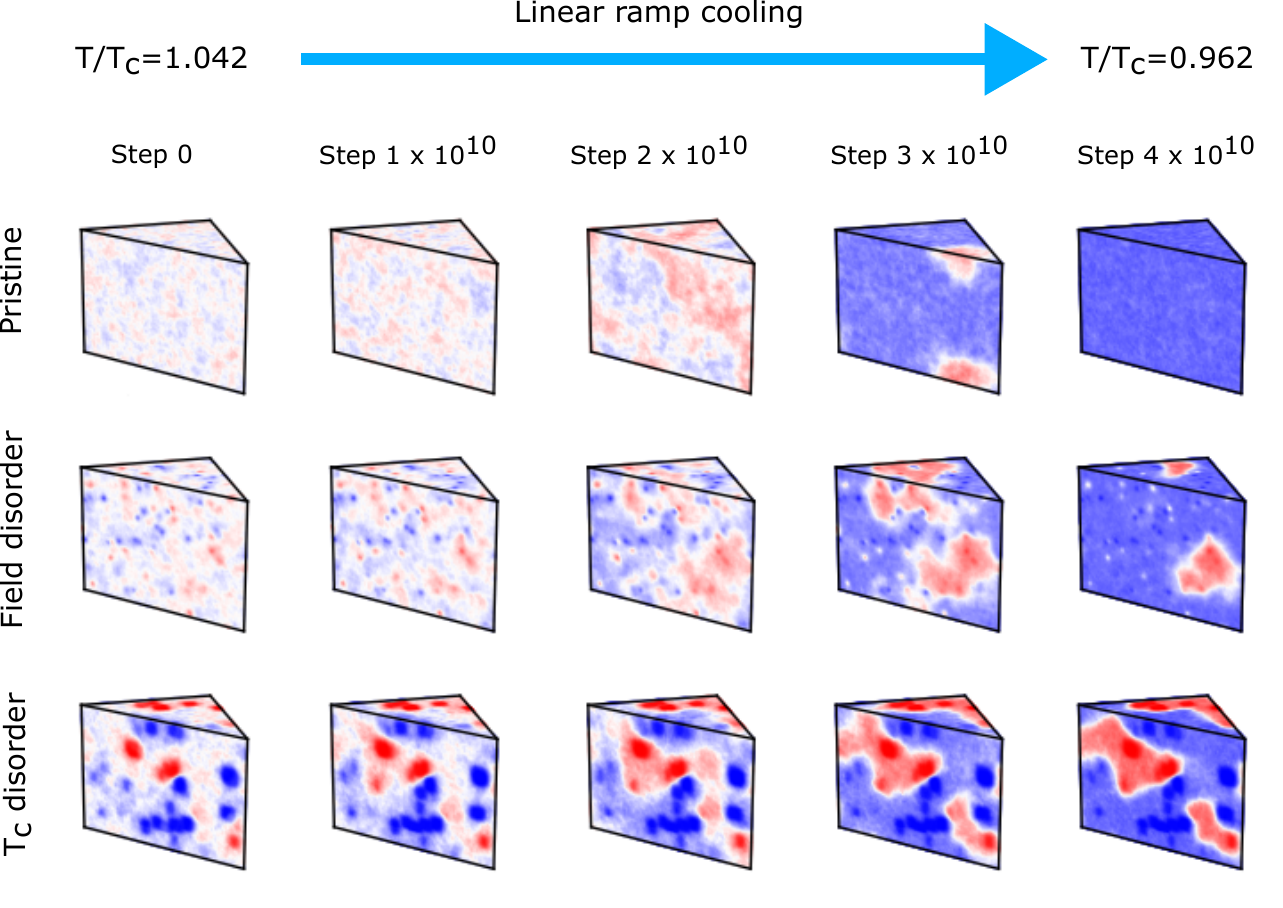}
    \caption{\textbf{Ising model metastable domain wall formation upon cooling.} The top row shows snapshots of the system state at the indicated step numbers for a pristine Ising simulation. The middle row is for field disorder and the bottom row shows $T_c$ disorder. All simulations are performed on a 100 $\times$ 100 $\times$ 100 grid with linear ramp cooling at a rate of $\Delta T/T_c = -2\times10^{-12}$ per step. In this figure, $T_c$ refers to the pristine system $T_c$ which is approximately 4.51 in dimensionless units. The simulation was initiated above $T_c$ at a temperature of $T/T_c = 1.042$. At this temperature, $10^{10}$ steps were performed to ensure equilibration. Then, the linear cooling was turned on. The step labels indicate steps after starting the linear cooling. The simulation was run until reaching a temperature of $T/T_c = 0.962$.}
    \label{fig:domain_wall}
\end{figure*}